%% file: magnetochronology-vidotto.tex
\documentclass[useAMS,usenatbib,usegraphicx]{mn2e}
\usepackage{amssymb,amsmath, color, url, soul, ulem}
\usepackage{epstopdf}
\defcitealias{1972ApJ...171..565S}{S72}

\defcitealias{Petit2014}          {1}
\defcitealias{2013arXiv1311.3374M}{2}
\defcitealias{2011A_A...530A..73C}{3}
\defcitealias{2006ARep...50..579K}{4}
\defcitealias{2013arXiv1310.7620D}{5}
\defcitealias{2010ApJ...714..384R}{6}
\defcitealias{2012ApJ...753...76W}{7}
\defcitealias{2008A_A...488..771J}{8}
\defcitealias{2003AJ....125.1980K}{9}
\defcitealias{2011ApJ...743...48W}{10}
\defcitealias{2009ApJ...698.1068P}{11}
\defcitealias{2007Ap.....50..187B}{12}
\defcitealias{2008MNRAS.388...80P}{13}
\defcitealias{2003A_A...397..147P}{14}
\defcitealias{2011AN....332..866M}{15}
\defcitealias{2012NewA...17..537X}{16}
\defcitealias{2012A_A...540A.138M}{17}
\defcitealias{2009IAUS..258..395G}{18}
\defcitealias{2001A_A...379..279Q}{19}
\defcitealias{2009A_A...501..941H}{20}
\defcitealias{2003A_A...397..987C}{21}
\defcitealias{2004A_A...417..651S}{22}
\defcitealias{2009A_A...508L...9P}{23}
\defcitealias{2002ApJ...574..412W}{24}
\defcitealias{2008ApJ...687.1264M}{25}
\defcitealias{2013A_A...556A..53E}{26}
\defcitealias{Folsom2014}         {27}
\defcitealias{2008hsf2.book..757T}{28}
\defcitealias{Waite2014}          {29}
\defcitealias{2011PASA...28..323W}{30}
\defcitealias{2011A_A...532A..10M}{31}
\defcitealias{2008ApJ...689.1127M}{32}
\defcitealias{1995A_A...302..775G}{33}
\defcitealias{1999MNRAS.302..437D}{34}
\defcitealias{2009A_ARv..17..251S}{35}
\defcitealias{2011MNRAS.410.2472A}{36}
\defcitealias{2013ApJ...766....6B}{37}
\defcitealias{2005ApJ...628L..69L}{38}
\defcitealias{2011MNRAS.413.1949W}{39}
\defcitealias{2011MNRAS.413.1922M}{40}
\defcitealias{2006MNRAS.370..468M}{41}
\defcitealias{2003A_A...411..595S}{42}
\defcitealias{2003A_A...399..983W}{43}
\defcitealias{1998ApJ...499L.199S}{44}
\defcitealias{2003A_A...410..671M}{45}
\defcitealias{2014MNRAS.438L..11B}{46}
\defcitealias{2003MNRAS.345.1145D}{47}
\defcitealias{2004A_A...417.1047K}{48}
\defcitealias{2004ApJ...614..386B}{49}
\defcitealias{2009MNRAS.398.1383F}{50}
\defcitealias{2013MNRAS.435.1451F}{51}
\defcitealias{2005A_A...443..609S}{52}
\defcitealias{2010A_A...515A..98P}{53}
\defcitealias{2008MNRAS.385.1179D}{54}
\defcitealias{2007MNRAS.374L..42C}{55}
\defcitealias{2008ApJ...687.1339K}{56}
\defcitealias{2003A_A...407..679U}{57}
\defcitealias{2006ApJ...648..683G}{58}
\defcitealias{2012MNRAS.423.1006F}{59}
\defcitealias{2010MNRAS.406..409F}{60}
\defcitealias{2006A_A...460..251M}{61}
\defcitealias{2008MNRAS.390..545D}{62}
\defcitealias{2014MNRAS.438.1162V}{63}
\defcitealias{2008MNRAS.390..567M}{64}
\defcitealias{2010MNRAS.407.2269M}{65}
\defcitealias{1995ApJ...450..392S}{66}
\defcitealias{2008MNRAS.384...77M}{67}
\defcitealias{2000ApJ...528..537P}{68}
\defcitealias{2010NatGe...3..637B}{69}
\defcitealias{2010MNRAS.409.1347D}{70}
\defcitealias{2014MNRAS.437.3202J}{71}
\defcitealias{2000A_A...358..593S}{72}
\defcitealias{2012ApJ...755...97G}{73}
\defcitealias{2007A_A...468..353G}{74}
\defcitealias{2008MNRAS.386.1234D}{75}
\defcitealias{2009MNRAS.398..189H}{76}
\defcitealias{2011AJ....141..127I}{77}
\defcitealias{1993ApJ...416..623F}{78}
\defcitealias{2013MNRAS.436..881D}{79}
\defcitealias{2012MNRAS.425.2948D}{80}
\defcitealias{2010A_A...519A.113G}{81}
\defcitealias{2011MNRAS.417..472D}{82}
\defcitealias{2011MNRAS.412.2454D}{83}
\defcitealias{2011A_A...530A...1A}{84}
\defcitealias{2010MNRAS.402.1426D}{85}
\defcitealias{2010A_A...519A..34P}{86}
\defcitealias{2011MNRAS.417.1747D}{87}
\defcitealias{2012ApJ...747..142S}{88}

\newcommand{\bv}{\langle |B_V| \rangle}
\newcommand{\bi}{\langle |B_I| \rangle}
\newcommand{\ro}{\rm Ro}
\newcommand{\Rx}{L_X/L_{\rm bol}}

\title{Stellar magnetism: empirical trends with age and rotation}
\author[A.~A.~Vidotto et al.]{A.~A.~Vidotto$^{1,2}$\thanks{E-mail: Aline.Vidotto@unige.ch}, {S.~G.~Gregory}$^1$, {M.~Jardine}$^{1}$, {J.~F.~Donati}$^{3}$, {P.~Petit}$^{3}$, {J.~Morin}$^{4}$, \newauthor  {C.~P.~Folsom}$^{3}$, {J.~Bouvier}$^{5}$, {A.~C.~Cameron}$^{1}$, {G.~Hussain}$^{6}$, {S.~Marsden}$^{7}$, {I.~A.~Waite}$^{7}$,   \newauthor  {R.~Fares}$^{1}$, {S. Jeffers}$^{8}$,  {J.~D.~do~Nascimento~Jr}$^{9,10}$
\\ 
$^{1}$SUPA, School of Physics and Astronomy, University of St Andrews, North Haugh, St Andrews, KY16 9SS, UK\\
$^{2}$Observatoire de Gen\'eve, Universit\'e de Gen\`eve, Chemin des Mailletes 51, Versoix, 1290, Switzerland\\
$^{3}$LATT - CNRS/Universit\'e de Toulouse, 14 Av.~E.~Belin, Toulouse, F-31400, France\\
$^{4}$LUPM-UMR5299, CNRS \& Universit\'e Montpellier II, Place E.~Bataillon, Montpellier, F-34095, France \\
$^{5}$UJF-Grenoble 1/CNRS-INSU, Institut de Plan\'etologie et d'Astrophysique de Grenoble (IPAG) UMR 5274, Grenoble, F-38041, France\\
$^{6}$ESO, Karl-Schwarzschild-Strasse 2, D-85748, Garching bei M\"nchen, Germany\\
$^{7}$Computational Engineering and Science Research Centre, University of Southern Queensland, Toowoomba, 4350, Australia\\
$^{8}$Institut f\"ur Astrophysik, Georg-August-Universit\"at, Friedrich-Hund-Platz 1, D-37077, Goettingen, Germany\\
$^{9}$Dep. de Fisica Te\'{o}rica e Exp., Un. Federal do Rio Grande do Norte, CEP: 59072-970 Natal, RN, Brazil\\
$^{10}$Harvard-Smithsonian Center for Astrophysics, 60 Garden Street, Cambridge, MA 02138, USA}
\begin{document}

\date{Accepted . Received ; in original form}

\pagerange{\pageref{firstpage}--\pageref{lastpage}} \pubyear{2014}

\maketitle

\label{firstpage}

%%%%%%%%%%%%%%%%%%%%%%%%%%%%%%%%%%%%%%%%%%%%%%%%
\begin{abstract}
We investigate how the observed large-scale surface magnetic fields of low-mass stars ($\sim$0.1 -- 2$\,{\rm M}_\odot$), reconstructed through Zeeman-Doppler imaging (ZDI), vary with age $t$, rotation and X-ray emission.  Our sample consists of $104$ magnetic maps of $73$ stars, from accreting pre-main sequence to main-sequence objects ($1 ~{\rm Myr} \lesssim t \lesssim 10$~Gyr). For non-accreting dwarfs we empirically find that the unsigned average large-scale surface field $\bv$ is related to age as {$t^{-0.655 \pm 0.045}$}. This relation has a similar dependence to that identified by Skumanich (1972), used as the basis for gyrochronology. Likewise, our relation could be used as an age-dating method (``magnetochronology''). The trends with rotation we find for the large-scale stellar magnetism are consistent with the trends found from Zeeman broadening measurements (sensitive to large- and small-scale fields). These similarities indicate that the fields recovered from both techniques are coupled to each other, suggesting that small- and large-scale fields could share the same dynamo field generation processes. For the accreting objects, fewer statistically significant relations are found, with one being a correlation between the unsigned magnetic flux $\Phi_V$ and $P_{\rm rot}$. We attribute this to a signature of star-disc interaction, rather than being driven by the dynamo.  
\end{abstract}
\begin{keywords}
stars: activity -- stars: evolution -- stars: magnetic field --  stars: rotation -- stars: planetary systems -- techniques: polarimetric 
\end{keywords}

%%%%%%%%%%%%%%%%%%%%%%%%%%%%%%%%%%%%%%%%%%%%%%%%%%
\section{INTRODUCTION}\label{sec.intro}
Magnetic fields play an important role in stellar evolution. For low-mass stars, the magnetic field is believed to regulate stellar rotation from the early stages of star formation until the ultimate stages of the life of a star. In their youngest phases, the stellar magnetic field lines interact with accretion discs to prevent what would have been a rapid spin-up of the star, caused by accretion of material with high angular momentum and also the stellar contraction \citep[e.g.,][]{2013arXiv1309.7851B}. After the accretion phase is over and the disc has dissipated, the contraction of the star towards the zero-age-main sequence (ZAMS) provides an abrupt spin up. From that phase onwards, `isolated' stars (single stars and stars in multiple systems with negligible tidal interaction, such as the ones adopted in our sample) slowly spin down as they age \citep[e.g.][]{2013A&A...556A..36G}. This fact was first observed by \citet[][S72, from now on]{1972ApJ...171..565S}, who empirically determined that the projected rotational velocities  $v \sin(i)$ of G-type stars in the main-sequence (MS) phase decrease with age $t$ as $v\sin(i) \propto t^{-1/2}$. This relation, called the ``Skumanich law'', serves as the basis of the gyrochoronology method \citep{2003ApJ...586..464B}, which yields age estimates based on rotation measurements. The rotational braking observed by \citetalias{1972ApJ...171..565S} is believed to be caused by stellar winds, which, outflowing along magnetic field lines, are able to efficiently remove the angular momentum of the star \citep[e.g.,][]{1958ApJ...128..664P,1962AnAp...25...18S,1967ApJ...148..217W}. 

Indicators of magnetic activity, such as surface spot coverage, emission from the chromosphere, transition region or corona, have been recognised to be closely linked to rotation \citep[e.g., S72;][]{1984A&A...133..117V,1984ApJ...279..763N,1997JGR...102.1641A,2007LRSP....4....3G,2012A&A...546A.117G,2012LRSP....9....1R}. However, the magnetic activity-rotation relation breaks for rapidly rotating stars, where the indicators of stellar magnetism saturate and become independent of  rotation. A saturation of the dynamo operating inside the star, inhibiting the increase of magnetism with rotation rate, has been attributed to explain the activity saturation observed in low-period stars \citep{1984A&A...133..117V}, but alternative explanations also exist \citep[e.g.,][]{1991ApJ...376..204M,1999A&A...346..883J,2007A&A...473..501A}. 

The average unsigned surface magnetic field $\langle |B_I| \rangle$, as measured by Zeeman-induced line broadening of unpolarised light (Stokes I), also correlates with rotation, in a similar way as the indicators of magnetic activity do (i.e., as one goes towards faster rotating stars, $\langle |B_I| \rangle$ increases until it reaches a saturation plateau; \citealt{2009ApJ...692..538R}). Because $\langle |B_I| \rangle$ is the product of the intensity-weighted surface filling factor of active regions $f$ and the mean unsigned field strength in the regions $B_I$ ($\langle |B_I| \rangle=fB_I$), it is still debatable whether the saturation occurs in the filling factor $f$ of magnetically active regions or in the stellar magnetism itself or in both \citep{1994ASPC...64..477S,1996IAUS..176..237S,2001ASPC..223..292S,2009ApJ...692..538R}. 

Although Zeeman broadening yields estimates of the average of the total (small and large scales) unsigned surface field strength,  it does not provide information on the magnetic topology \citep{2013AN....334...48M}. For that, a complementary magnetic field characterisation technique, namely Zeeman-Doppler Imaging \citep[ZDI, e.g.,][]{1997A&A...326.1135D}, should be employed. The ZDI technique consists of analysing a series of circularly polarised spectra (Stokes V signatures) to recover information about the large-scale magnetic field (its intensity and orientation). In this work, we take advantage of the increasing number of stars with surface magnetic fields mapped through the ZDI technique and investigate how their large-scale surface magnetism varies with age, rotation and X-ray luminosity (an activity index). In the past decade,  ZDI  has been used to reconstruct the topology and intensity of the surface magnetic fields of roughly one hundred stars \citep[for a recent review of the survey, see][]{2009ARA&A..47..333D}. Since the ZDI technique measures the magnetic flux averaged over surface elements, regions of opposite magnetic polarity within the element resolution cancel each other out \citep{2010MNRAS.404..101J,2011MNRAS.410.2472A}. As a consequence, the ZDI magnetic maps are limited to measuring large-scale magnetic field. 

Because the small-scale field decays faster with height above the stellar surface than the large-scale field \citep[e.g.,][]{2014MNRAS.439.2122L}, only the latter permeates the stellar wind. If indeed magnetised stellar winds are the main mechanism of removing angular momentum from the star in the MS phase, one should expect the large-scale field to correlate with rotation and age. Likewise, a correlation between rotation and magnetism should also be expected if rotation is the driver of stellar magnetism through dynamo field generation processes. The interaction between magnetism, rotation and age is certainly complex and empirical relations, such as the ones derived in this work, provide important constraints for studies of rotational evolution and stellar dynamos. 

This paper is organised as follows. We present our sample of stars in Section~\ref{sec.sample}. Section~\ref{sec.relations} shows the empirically-derived trends with magnetism we find within our data. In Section~\ref{sec.discussion}, we discuss how the results obtained using the Zeeman broadening technique compare to the ones derived from Zeeman-Doppler Imaging (Section~\ref{sec.comparison}), we investigate the presence of saturation in the large-scale field  (\ref{sec.saturation}),  analyse whether stars hosting hot-Jupiters present different magnetism compared to stars lacking hot-Jupiters (\ref{sec.hjhosts}) and discuss the trends obtained for the pre-main sequence (PMS) accreting stars (\ref{sec.accreting}). In Section~\ref{sec.magnetochronology}, we discuss the impact of our findings as a new way to assess stellar ages and as a valuable observational input for dynamo studies and stellar mass loss evolution. Our summary and conclusions are presented in Section~\ref{sec.conclusions}.

%%%%%%%%%%%%%%%%%%%%%%%%%%%%%%
\section{The sample of stars}\label{sec.sample}
The stars considered in this study consist of $73$ late-F, G, K, and M dwarf stars, in the PMS to MS phases. All have had their large-scale surface magnetic fields reconstructed using the ZDI technique, with some having been observed at multiple epochs, as listed in Table~\ref{table}. The magnetic maps, $104$ in total, have either been published elsewhere \citep{1999MNRAS.302..437D,2003MNRAS.345.1145D,2008MNRAS.390..545D,2008MNRAS.385.1179D,2008MNRAS.386.1234D,2010MNRAS.409.1347D,2010MNRAS.402.1426D,2011MNRAS.412.2454D,2011MNRAS.417..472D,2011MNRAS.417.1747D,2012MNRAS.425.2948D,2013MNRAS.436..881D, 2006MNRAS.370..468M,2011MNRAS.413.1922M,2007MNRAS.374L..42C,2008MNRAS.384...77M,2008MNRAS.390..567M,2010MNRAS.407.2269M,2008MNRAS.388...80P,2009A_A...508L...9P,2009MNRAS.398..189H,2009MNRAS.398.1383F,2010MNRAS.406..409F,2012MNRAS.423.1006F,2013MNRAS.435.1451F,2011AN....332..866M,2012A_A...540A.138M,2011MNRAS.413.1949W,2013arXiv1310.7620D} or are in process of being published (Folsom et al.~2014 in prep, Petit et al.~2014 in prep, Waite et al.~2014 in prep). Although the reconstructed maps provide the distribution of magnetic fields at the stellar surface, in this paper we only use the unsigned average field strengths $\bv$ (i.e., integrated over the surface of the star).\footnote{In order to differentiate between field strengths derived from Stokes V measurements (ZDI) and from Stokes I (Zeeman broadening), we use the indices $V$ and $I$, respectively.} In the present work, $\bv$ is calculated based on the radial component of the observed surface field, as we are mainly interested in the field associated with the stellar wind \citep{2013MNRAS.431..528J}. We also consider the Sun in our dataset. For the solar magnetic field, we use the magnetograms from NSO - Kitt Peak data archive at solar maximum and minimum (Carrington rotations CR1851 and CR1907, respectively). To allow a direct comparison of the solar and stellar magnetic fields, we restrict the reconstruction of the solar surface fields to a maximum order of $l_{\rm max} =3$ of the spherical harmonic expansion (note, for instance, that modes with $l \lesssim 3$ already contain the bulk of the  total photospheric magnetic energy in solar-type stars, \citealt{2008MNRAS.388...80P}). 

Table~\ref{table} lists the general characteristics of the stars considered here, including quantities such as age $t$ (whenever available), rotation period $P_{\rm rot}$, $\langle |B_V| \rangle$, Rossby number Ro, X-ray luminosity $L_X$ and $L_X/L_{\rm bol}$, where $L_{\rm bol}$ is the bolometric luminosity. The measurement errors associated with these quantities are described in Appendix~\ref{ap.sample}. Among the $73$ stars in our sample, $61$ objects have age estimates (totaling $90$ maps), which were collected from the literature and are based on different methods. For the PMS accreting stars, ages were derived using the stellar evolution models of \citet{2000A_A...358..593S}, as derived in \citet{2012ApJ...755...97G} and \citet{2013MNRAS.436..881D}. For the remaining stars, methods used for deriving ages include, for example, isochrones, lithium abundance, kinematic convergent point, gyrochronology, chromospheric activity. Note also that some of the stars in our sample are members of associations and open clusters and have, therefore, a reasonably well-constrained age (often derived with multiple methods). The last column of Table~\ref{table} lists the references for all the values adopted in this paper. In particular, the references from which ages were obtained are presented in boldface. 

\begin{table*} 
\centering
\caption{The objects in our sample. Columns are: star name, spectral type, mass, radius, rotation period, Rossby number, age, X-ray luminosity, X-ray-to-bolometric luminosity ratio, average large-scale unsigned surface magnetic field and its observation epoch (year and month). The measurement errors associated with these quantities are described in Appendix~\ref{ap.sample}. References for the values compiled in this table are shown in the last column. In boldface are the references from which the ages adopted in this paper were obtained. \label{table}}    
\begin{tabular}{llccccccccccccccccc}  
\hline
Star	&	Sp.		&	$	M_\star 	$	&	$	R_\star		$	&	$	P_{\rm rot}		$	&	$	\ro		$	&	age & $\log\left[\frac{L_X}{\rm erg/s}\right]$ &$\log\left[\frac{L_X}{L_{\rm bol}}\right]$& $	\bv		$	 &  Obs.& Ref. \\ 
ID     &	Type &	$(M_\odot)	$	&	$(R_\odot)	$	&(d)&  & (Myr)& & &$	(G)	$ & epoch&  & 
\input{latex_table_a.tex}	
\hline
\end{tabular}
\\

\citetalias{Petit2014}          : Petit et al.~(2014) in prep.;
\citetalias{2013arXiv1311.3374M}: \citet{2013arXiv1311.3374M};
\citetalias{2011A_A...530A..73C}: \citet{2011A_A...530A..73C};
\citetalias{2006ARep...50..579K}: \citet{2006ARep...50..579K};
\citetalias{2013arXiv1310.7620D}: \citet{2013arXiv1310.7620D};
\citetalias{2010ApJ...714..384R}: \citet{2010ApJ...714..384R};
\citetalias{2012ApJ...753...76W}: \citet{2012ApJ...753...76W};
\citetalias{2008A_A...488..771J}: \citet{2008A_A...488..771J};
\citetalias{2003AJ....125.1980K}: \citet{2003AJ....125.1980K};
\citetalias{2011ApJ...743...48W}: \citet{2011ApJ...743...48W};
\citetalias{2009ApJ...698.1068P}: \citet{2009ApJ...698.1068P};
\citetalias{2007Ap.....50..187B}: \citet{2007Ap.....50..187B};
\citetalias{2008MNRAS.388...80P}: \citet{2008MNRAS.388...80P};
\citetalias{2003A_A...397..147P}: \citet{2003A_A...397..147P};
\citetalias{2011AN....332..866M}: \citet{2011AN....332..866M};
\citetalias{2012NewA...17..537X}: \citet{2012NewA...17..537X};
\citetalias{2012A_A...540A.138M}: \citet{2012A_A...540A.138M};
\citetalias{2009IAUS..258..395G}: \citet{2009IAUS..258..395G};
\citetalias{2001A_A...379..279Q}: \citet{2001A_A...379..279Q};
\citetalias{2009A_A...501..941H}: \citet{2009A_A...501..941H};
\citetalias{2003A_A...397..987C}: \citet{2003A_A...397..987C};
\citetalias{2004A_A...417..651S}: \citet{2004A_A...417..651S};
\citetalias{2009A_A...508L...9P}: \citet{2009A_A...508L...9P};
\citetalias{2002ApJ...574..412W}: \citet{2002ApJ...574..412W};
\citetalias{2008ApJ...687.1264M}: \citet{2008ApJ...687.1264M};
\citetalias{2013A_A...556A..53E}: \citet{2013A_A...556A..53E};
\citetalias{Folsom2014}         : Folsom et al.~(2014) in prep.;
\citetalias{2008hsf2.book..757T}: \citet{2008hsf2.book..757T};
\citetalias{Waite2014}          : Waite et al.~(2014) in prep.;
\citetalias{2011PASA...28..323W}: \citet{2011PASA...28..323W};
\citetalias{2011A_A...532A..10M}: \citet{2011A_A...532A..10M};
\citetalias{2008ApJ...689.1127M}: \citet{2008ApJ...689.1127M}
\end{table*}
\begin{table*} 
\centering
\contcaption{The objects in our sample.}  
\begin{tabular}{llccccccccccccccccc}  
\hline
Star	&	Sp.		&	$	M_\star 	$	&	$	R_\star		$	&	$	P_{\rm rot}		$	&	$	\ro		$	&	age & $\log\left[\frac{L_X}{\rm erg/s}\right]$ &$\log\left[\frac{L_X}{L_{\rm bol}}\right]$& $	\bv		$	 &  Obs.& Ref. \\ 
ID     &	Type &	$(M_\odot)	$	&	$(R_\odot)	$	&(d)&  & (Myr)& & &$	(G)	$ & epoch&  &  
\input{latex_table_b.tex}
\hline 
\end{tabular}
\\ 

\citetalias{1995A_A...302..775G}: \citet{1995A_A...302..775G};
\citetalias{1999MNRAS.302..437D}: \citet{1999MNRAS.302..437D};
\citetalias{2009A_ARv..17..251S}: \citet{2009A_ARv..17..251S};
\citetalias{2011MNRAS.410.2472A}: \citet{2011MNRAS.410.2472A};
\citetalias{2013ApJ...766....6B}: \citet{2013ApJ...766....6B};
\citetalias{2005ApJ...628L..69L}: \citet{2005ApJ...628L..69L};
\citetalias{2011MNRAS.413.1949W}: \citet{2011MNRAS.413.1949W};
\citetalias{2011MNRAS.413.1922M}: \citet{2011MNRAS.413.1922M};
\citetalias{2006MNRAS.370..468M}: \citet{2006MNRAS.370..468M};
\citetalias{2003A_A...411..595S}: \citet{2003A_A...411..595S};
\citetalias{2003A_A...399..983W}: \citet{2003A_A...399..983W};
\citetalias{1998ApJ...499L.199S}: \citet{1998ApJ...499L.199S};
\citetalias{2003A_A...410..671M}: \citet{2003A_A...410..671M};
\citetalias{2014MNRAS.438L..11B}: \citet{2014MNRAS.438L..11B};
\citetalias{2003MNRAS.345.1145D}: \citet{2003MNRAS.345.1145D};
\citetalias{2004A_A...417.1047K}: \citet{2004A_A...417.1047K};
\citetalias{2004ApJ...614..386B}: \citet{2004ApJ...614..386B};
\citetalias{2009MNRAS.398.1383F}: \citet{2009MNRAS.398.1383F};
\citetalias{2013MNRAS.435.1451F}: \citet{2013MNRAS.435.1451F};
\citetalias{2005A_A...443..609S}: \citet{2005A_A...443..609S};
\citetalias{2010A_A...515A..98P}: \citet{2010A_A...515A..98P};
\citetalias{2008MNRAS.385.1179D}: \citet{2008MNRAS.385.1179D};
\citetalias{2007MNRAS.374L..42C}: \citet{2007MNRAS.374L..42C};
\citetalias{2008ApJ...687.1339K}: \citet{2008ApJ...687.1339K};
\citetalias{2003A_A...407..679U}: \citet{2003A_A...407..679U};
\citetalias{2006ApJ...648..683G}: \citet{2006ApJ...648..683G};
\citetalias{2012MNRAS.423.1006F}: \citet{2012MNRAS.423.1006F};
\citetalias{2010MNRAS.406..409F}: \citet{2010MNRAS.406..409F};
\citetalias{2006A_A...460..251M}: \citet{2006A_A...460..251M};
\citetalias{2008MNRAS.390..545D}: \citet{2008MNRAS.390..545D};
\citetalias{2014MNRAS.438.1162V}: \citet{2014MNRAS.438.1162V};
\citetalias{2008MNRAS.390..567M}: \citet{2008MNRAS.390..567M};
\citetalias{2010MNRAS.407.2269M}: \citet{2010MNRAS.407.2269M};
\citetalias{1995ApJ...450..392S}: \citet{1995ApJ...450..392S};
\citetalias{2008MNRAS.384...77M}: \citet{2008MNRAS.384...77M};
\citetalias{2000ApJ...528..537P}: \citet{2000ApJ...528..537P};
\citetalias{2010NatGe...3..637B}: \citet{2010NatGe...3..637B};
\citetalias{2010MNRAS.409.1347D}: \citet{2010MNRAS.409.1347D};
\citetalias{2014MNRAS.437.3202J}: \citet{2014MNRAS.437.3202J};
\citetalias{2000A_A...358..593S}: \citet{2000A_A...358..593S};
\citetalias{2012ApJ...755...97G}: \citet{2012ApJ...755...97G};
\citetalias{2007A_A...468..353G}: \citet{2007A_A...468..353G};
\citetalias{2008MNRAS.386.1234D}: \citet{2008MNRAS.386.1234D};
\citetalias{2009MNRAS.398..189H}: \citet{2009MNRAS.398..189H};
\citetalias{2011AJ....141..127I}: \citet{2011AJ....141..127I};
\citetalias{1993ApJ...416..623F}: \citet{1993ApJ...416..623F};
\citetalias{2013MNRAS.436..881D}: \citet{2013MNRAS.436..881D};
\citetalias{2012MNRAS.425.2948D}: \citet{2012MNRAS.425.2948D};
\citetalias{2010A_A...519A.113G}: \citet{2010A_A...519A.113G};
\citetalias{2011MNRAS.417..472D}: \citet{2011MNRAS.417..472D};
\citetalias{2011MNRAS.412.2454D}: \citet{2011MNRAS.412.2454D};
\citetalias{2011A_A...530A...1A}: \citet{2011A_A...530A...1A};
\citetalias{2010MNRAS.402.1426D}: \citet{2010MNRAS.402.1426D};
\citetalias{2010A_A...519A..34P}: \citet{2010A_A...519A..34P};
\citetalias{2011MNRAS.417.1747D}: \citet{2011MNRAS.417.1747D};
\citetalias{2012ApJ...747..142S}: \citet{2012ApJ...747..142S}

\end{table*}

In order to investigate possible correlations in our data, we perform power-law fits of any two quantities $x=\log(X)$ and $y=\log(Y)$ (fitted through linear least-squares fits to logarithms), such that $y = q +p x$ (or $Y=10^q X^p)$, with $q$ and $p$ being the coefficients derived in the linear regression. These regressions were obtained using the bisector ordinary least-squares method \citep{1990ApJ...364..104I}, which treats the $x$ and $y$ variables symmetrically  \citep{basu}. We opted such a fitting method because, for the quantities analysed here, the functional dependences of $x$ and $y$ are not clear. 

\begin{figure} 
\includegraphics[width=80mm]{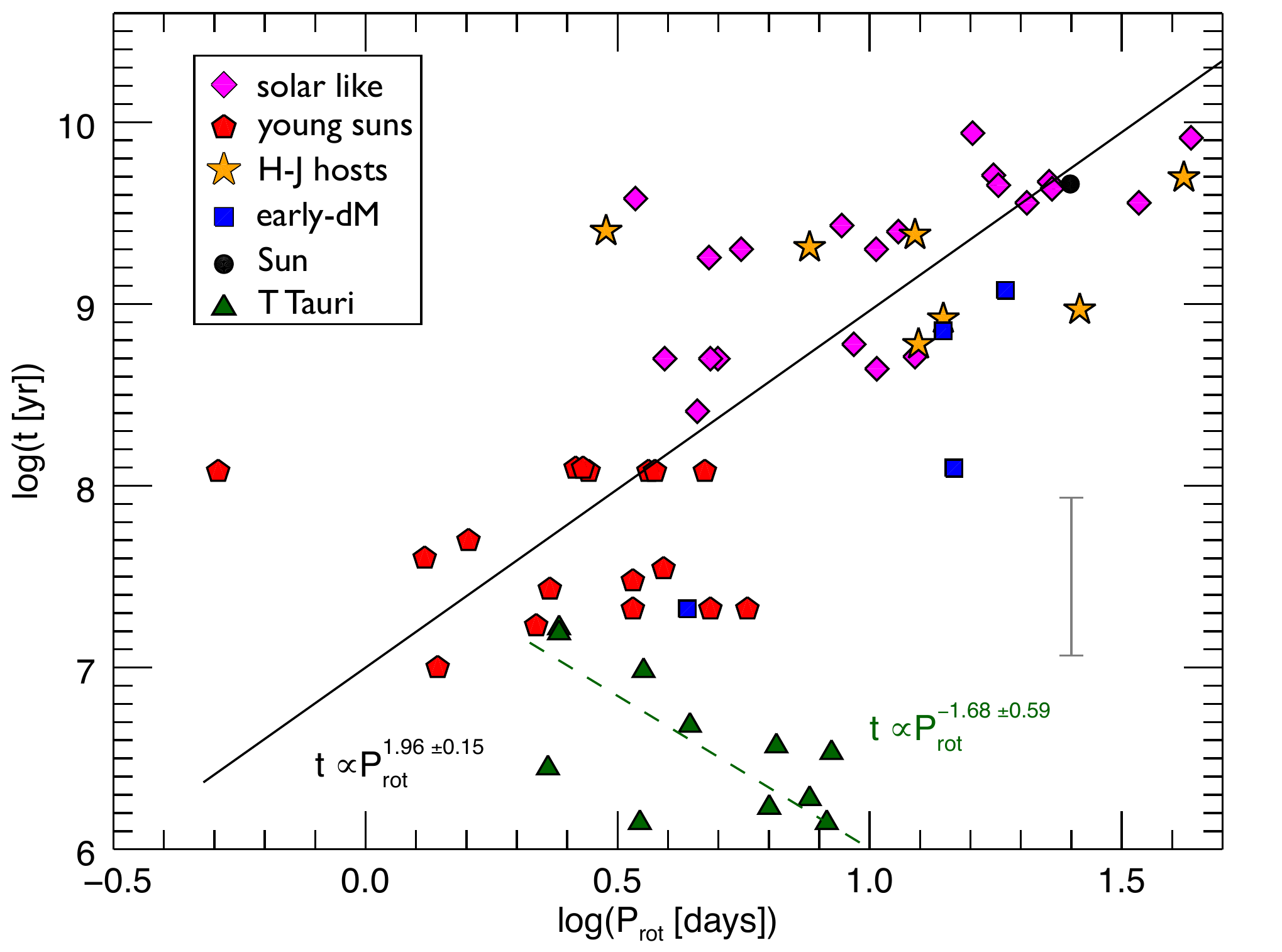}%5_age_prot}
\caption{Correlation between age $t$ and rotation period $P_{\rm rot}$ for the stars in our sample, indicating that the non-accreting stars follow the Skumanich law ($t \propto P_{\rm rot}^{2}$). The solid (dashed) line is a power-law fit to our sample of non-accreting (accreting) objects. A typical error bar is indicated in grey (also in Figures~\ref{fig.B_age} to \ref{fig.ap2}).\label{fig.prot_age}}
\end{figure}

Before we present the analyses of the trends with magnetism, it is useful to compare how our data relate to the Skumanich law, where rotation period $P_{\rm rot}$ is related to age as $P_{\rm rot} \propto t^{1/2}$ or $t \propto P_{\rm rot}^2 $ (see~Figure~\ref{fig.prot_age}). The  power-law indices $p$ obtained for the non-accreting (solid line) and accreting (dashed line) stars are shown in Table~\ref{tab.powerlaw}, along with the Spearman's rank correlation coefficient $\rho$ and its probability under the null hypothesis (i.e., uncorrelated quantities). For the non-accreting stars, we find that {$t \propto P_{\rm rot}^{1.96 \pm 0.15 }$ ($\rho=0.76$)}, which is consistent with the Skumanich law. Note that the accreting stars (green points) follow a different behaviour to the remaining objects in our sample and, because of that, we treat them as a different population throughout this paper. The physics of accreting stars is more complex than that of the disc-less stars, as the former interact with their accretion discs through stellar magnetic field lines that thread their discs (for a recent review see \citealt{2013arXiv1309.7851B} and references therein). As a consequence, the presence of the disc controls the rotation of these stars \citep{2007ApJ...671..605C}. In addition, the young PMS stars will continue to contract once the disc has dispersed and, consequently, will spin up, while evolving towards the ZAMS. Because not enough time has passed since their formation from the gravitational collapse of their natal molecular clouds, they still have imprinted on them the initial conditions of their rotation and, therefore, possess a large spread in the $P_{\rm rot}$-$t$ diagram.

\begin{table*} 
\centering
\caption{Power-law indices ($Y\propto X^p$) computed using the bisector linear least-squares method, fitted to logarithms. The Spearman's rank correlation coefficient $\rho$ and its probability under the null hypothesis are also shown. Fits considering only non-accreting F, G, K and early-M dwarf stars, only PMS accreting stars and all the data in our sample are shown separately. %\hll{updated 10/feb}
\label{tab.powerlaw}}    
\begin{tabular}{llrrcrrcrrcccccc}  
\hline
&&\multicolumn{3}{c}{Fits for dwarf stars only} &\multicolumn{3}{c}{Fits for accreting stars only} &\multicolumn{3}{c}{Fits considering all the sample}  \\ \hline
$Y$ & $X$ & $\rho$ & Prob. & $p$ & $\rho$  & Prob.  &  $p$ & $\rho$ & Prob. & $p$\\
&&& (\%)&&&(\%)&&&(\%) \\
\hline
\input{fit_bisector_table_for_paper.tex}	
\hline
\end{tabular}
\\$^a$ Fits considering only points with $\ro \gtrsim 0.1$ (cf.~\S\ref{sec.comparison}).
\end{table*}

%%%%%%%%%%%%%%%%%%%%%%%%%%%%%%%%%%%%%
\section{Trends with magnetism}\label{sec.relations}
In this Section, we investigate possible trends between the following quantities:  $\langle |B_V| \rangle$,  $t$, $P_{\rm rot}$,  \ro , $L_X$, $L_X/L_{\rm bol}$ and unsigned magnetic flux $\Phi_V$. Table \ref{tab.powerlaw} summarises the results of our fits. {It is worth noting that, when analysed individually, each subset of objects (as presented in Table~\ref{table}) do not show correlations with high statistical significance due to their narrow range of parameters (e.g., ages, rotation periods). However, trends are more robust when the different subsets are combined together and the dynamic range increases.} For the non-accreting stars, all the relations have high statistical significance, with usually large Spearman's rank correlation coefficients and low probabilities of there not being correlations ($< 0.01$ \%). On the other hand, the relations we derive for the accreting stars are significantly poorer, with $|\rho | \lesssim 0.6$ and usually high probabilities of these quantities not being correlated, except for $\Phi_V$ versus $P_{\rm rot}$. The poorer fits are a result of the narrower range of parameters of this subset and also due to its relatively small number of data points ($11$ stars and $16$ magnetic maps). These objects will be discussed in more detail in Section~\ref{sec.ttauri}.  Next, we discuss a few selected trends for the non-accreting population.

%%%%%%%%%%%%%%%%%%%%%%%%%%%%%%%%%%%%%%%%
\subsection{Non-accreting stars}
%%%%%%%%%%%%%%%%%%%%%%%%%%%%%%%%%%%%%%%%%%%%%%%%%%%%%%%%%
\subsubsection{Correlation with age}
\begin{figure}
\includegraphics[width=80mm]{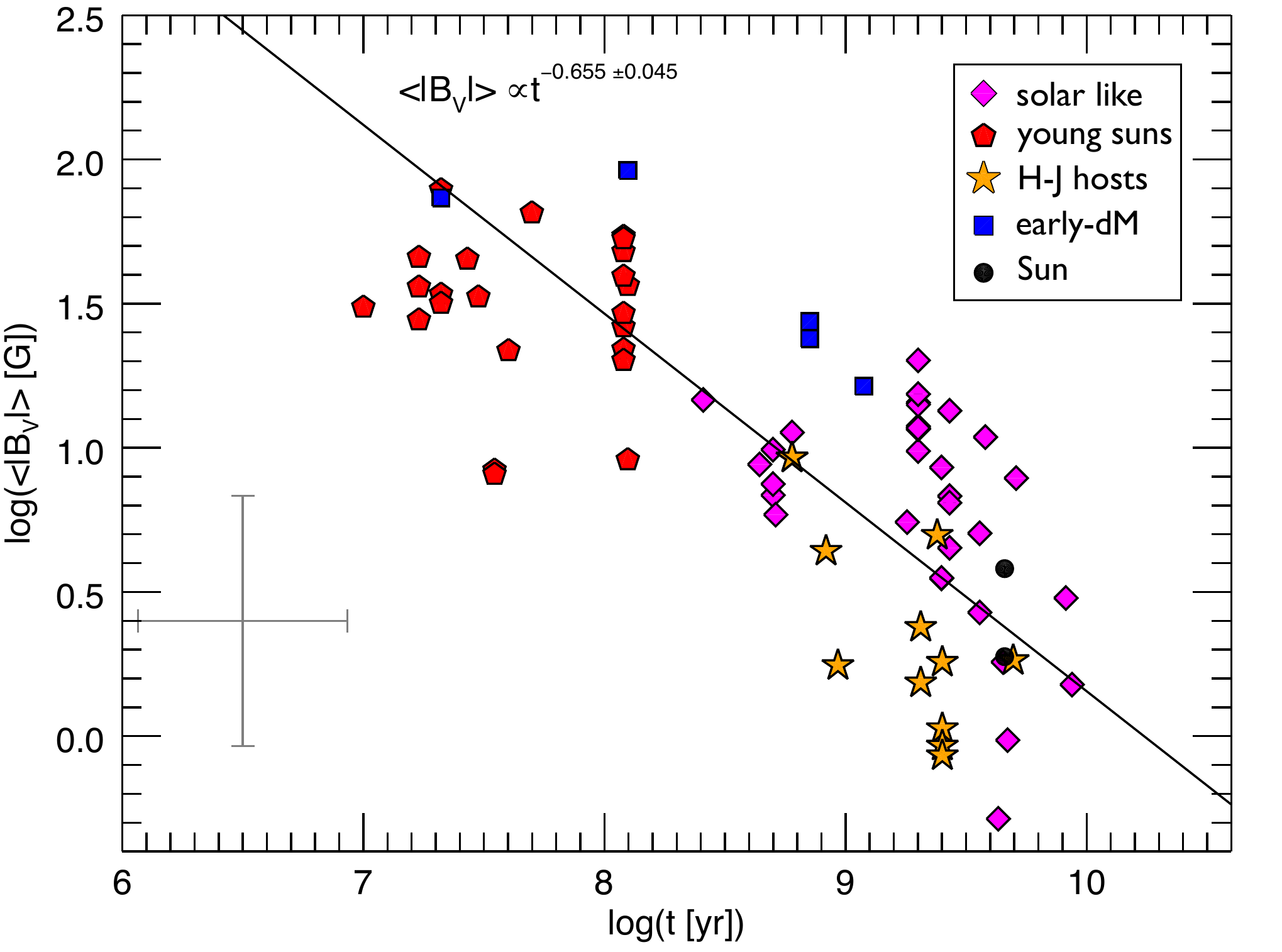}%5_B_age}\\
\caption{Correlation between the average large-scale field strength derived from the ZDI technique $\bv$ and age $t$, for the non-accreting stars in our sample. The trend found (solid line) has a similar age dependence as the Skumanich law ($\Omega_\star \propto t^{-0.5}$). This relation could be used as an alternative method to estimate the age of stars (``magnetochronology''). \label{fig.B_age}}
\end{figure}

In his seminal paper, \citetalias{1972ApJ...171..565S} predicted that magnetic fields decay as the inverse square of age, based on the age-rotation relation and further assuming that surface fields have a linear dependence with the rotation of the star (cf.~Section~\ref{subsec.prot-B}). In order to test this prediction, we show in Figure~\ref{fig.B_age} the trend we find between $\langle |B_V| \rangle$ and $t$ for the stars in our sample. The correlation we found holds for more than two orders of magnitude in $\langle |B_V| \rangle$ and three orders of magnitude in $t$ for the non-accreting stars. From our power-law fit (solid line), we find that $\langle |B_V| \rangle \propto t^{-0.655 \pm 0.045}$, which has a similar age-dependence as the Skumanich law ($\Omega_\star \propto t^{-0.5}$) and supports the magnetism-age prediction inferred by \citetalias{1972ApJ...171..565S} that there is magnetic field decay as the inverse square-root of age. A similar power-law dependence is found between the unsigned surface flux $\Phi_V= \bv 4 \pi R_\star^2$ and age ($\Phi_V \propto t^{-0.622\pm 0.042}$). 

%%%%%%%%%%%%%%%%%%%
%%%%%%%%%%%%%%%%%%%

\subsubsection{Correlation with rotation period}\label{subsec.prot-B}
\begin{figure}
\includegraphics[width=80mm]{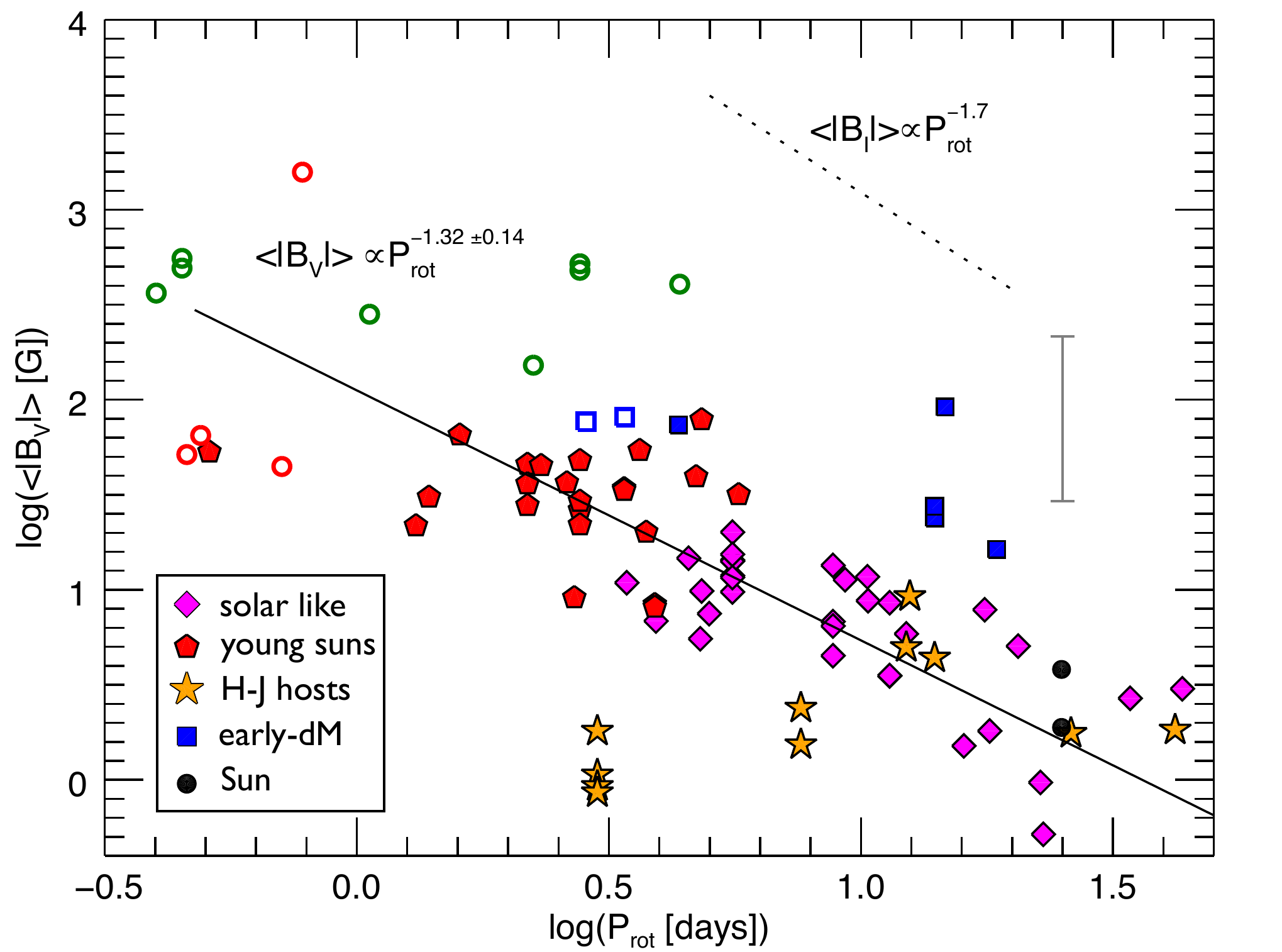}%5_B_prot}
\caption{Correlation between the average large-scale field strength derived from the ZDI technique $\bv$ and rotation period $P_{\rm rot}$, for the non-accreting stars in our sample. Our data support the presence of a linear-type dynamo for the large-scale field (i.e., $\bv \propto \Omega_\star \propto P_{\rm rot}^{-1}$) within {$3\sigma$}, although a large scatter exists. The open symbols (not considered in the fit) are saturated M dwarf stars without age estimates: blue squares for $M_\star \geq 0.4~M_\odot$ (early Ms), green circles for $0.2 < M_\star/M_\odot < 0.4$ (mid Ms) and red circles for $M_\star \leq 0.2~M_\odot$ (late Ms). The dotted line, at an arbitrary vertical offset, is indicative of the slope found from ZB measurements between $\bi$ and $P_{\rm rot}$ \citep{1996IAUS..176..237S}.
\label{fig.b_prot}}
\end{figure}

Stellar winds are believed to regulate the rotation of MS stars. The empirical Skumanich law, for example, can be theoretically explained using a simplified stellar wind model \citep{1967ApJ...148..217W}, if one assumes that the stellar magnetic field scales linearly with the rotation rate of the star $\Omega_\star$. To investigate whether our data support the presence of such a linear-type dynamo  ($B \propto \Omega_\star \propto P_{\rm rot}^{-1}$), we present how $\langle |B_V| \rangle$ scales with $P_{\rm rot}$  in Figure~\ref{fig.b_prot}. Our results show that {$\langle |B_V| \rangle \propto P_{\rm rot}^{-1.32 \pm 0.14}$ ($|\rho|=0.54$)}, indicating that our data  supports a linear-type dynamo of the large-scale field within {$3\sigma$}. A similar nearly linear trend is found between the unsigned surface flux $\Phi_V$ and $P_{\rm rot}$, with a larger correlation coefficient {$|\rho|=0.72$}. 

Although the correlation between $\langle |B_V| \rangle$ and $P_{\rm rot}$ indeed exists (with a negligible null-probability), this relation has a significant spread. One possible explanation for this spread could be that in the Weber-Davis theory of stellar winds, a very simplistic field geometry is assumed (a split monopole) with the entire surface of the star contributing to wind launching. However, the complexity of the magnetic field topology can play an important role in the rotational evolution of the star \citep[e.g.,][]{2009ApJ...703.1734V,2012MNRAS.423.3285V,2010ApJ...721...80C}. ZDI observations have shown that stellar magnetic field topologies can be much more complex than that of a split monopole. In addition, numerical simulations of stellar winds show that part of the large-scale surface field should consist of closed field lines, which do not contribute to angular momentum removal \citep[e.g.,][]{2014MNRAS.438.1162V}. The large spread in the $\langle |B_V| \rangle$-$P_{\rm rot}$ relation could therefore be explained by the differences in magnetic field topologies present in the stars of our sample.

%%%%%%%%%%%%%%%%%%%
%%%%%%%%%%%%%%%%%%%

%%%%%%%%%%%%%%%%%%%%%%%%%%%%%%%%%%%%%%%%%%
\subsubsection{Correlation with Rossby number}
Another possibility for the spread found in the relation between $\langle |B_V| \rangle$ and $P_{\rm rot}$ can be due to the fact that we are considering a broad range of spectral types. Traditionally, the use of Rossby number (\ro) instead of $P_{\rm rot}$ allows comparison across different spectral types, reducing the spread commonly noticed in trends involving $P_{\rm rot}$. \ro\ is defined as the ratio between $P_{\rm rot}$ and convective turnover time $\tau_c$. To calculate \ro\ for the non-accreting stars, we used the theoretical determinations of $\tau_c$ from \citet{2010A&A...510A..46L}. Appendix \ref{ap.rossby} shows how our results vary if we adopt different approaches for the calculation of $\tau_c$. For the {$8$} stars that have masses outside the mass interval  for which $\tau_c$ was computed  in \citet[][$0.6 \leq M_\star/M_\odot \leq 1.2$]{2010A&A...510A..46L}, we adopt the following approximation. Stars with a given age t and mass $M_\star \leq 0.6M_\odot$ were assumed to have $\tau_c=\tau_c(M_\star=0.6M_\odot, t)$ and for  $M_\star \geq 1.2 M_\odot$ were assumed to have $\tau_c=\tau_c(M_\star=1.2 M_\odot, t)$. As a result, for the former (latter) group, the calculated $\tau_c$ is a lower (upper) limit, while \ro\ is an upper (lower) limit. In this work, we do not assign errors to Rossby numbers, but we note that these values are model-dependent. For the accreting stars, \ro\ was derived from an update to the models of \citet{1996ApJ...457..340K}, as detailed by \citet{2012ApJ...755...97G}.  

\begin{figure}
\includegraphics[width=80mm]{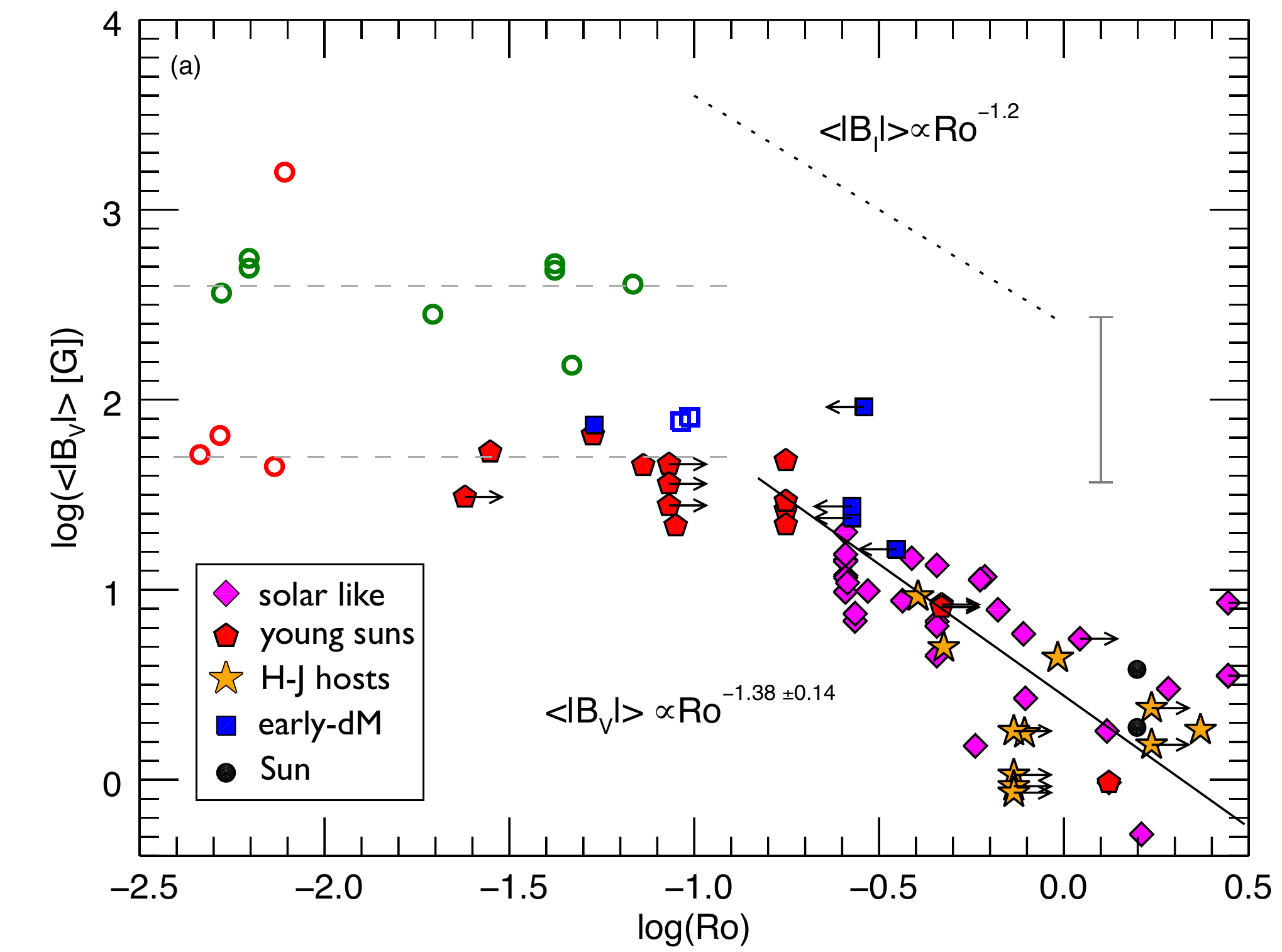}\\%5_B_rossby_landin}\\
\includegraphics[width=80mm]{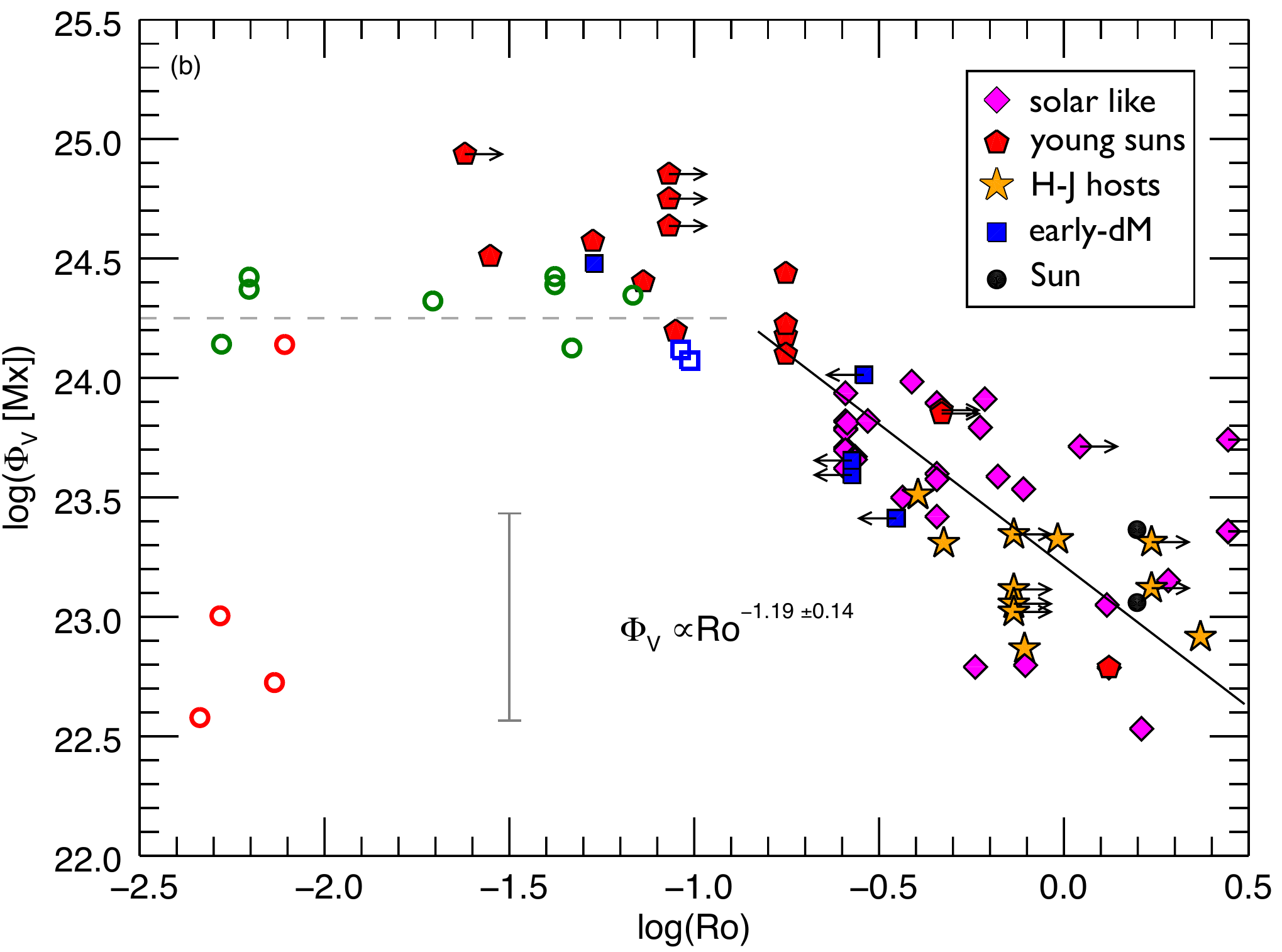}%5_flux_rossby_landin}
\caption{(a) Correlation between the average large-scale field strength derived from the ZDI technique $\bv$ and Rossby number \ro, for the non-accreting stars in our sample. Using Stokes I data, \citet{2009ApJ...692..538R} showed that  $\bi$ saturates for $\ro \lesssim 0.1$. \citet{2008MNRAS.390..545D} suggested that there might be two different levels of saturation (dashed lines) among the low-mass stars, caused by different efficiencies at producing large- and small-scale fields. (b) Same as in (a), but now considering the magnetic flux $\Phi_V$. Note that the bi-modality in the saturation level is removed if $\Phi_V$ is considered instead of $\bv$. Open symbols are as in Figure \ref{fig.b_prot}. Solid lines show power-law fits considering objects with $\ro\gtrsim 0.1$. The dotted line (arbitrary vertical offset) in the upper panel is indicative of the slope found from ZB measurements between $\bi$ and \ro\ \citep{2001ASPC..223..292S}. \label{fig.b_rossby}}
\end{figure}

In general, all our fits against \ro\ have larger unsigned Spearman's rank correlation coefficients than fits against $P_{\rm rot}$. Figure~\ref{fig.b_rossby}a shows $\bv$ as a function of \ro, where we find that {$\bv \propto \ro^{-1.38\pm0.14}$}. This relation will be further discussed later on Section~\ref{sec.comparison}.  Additionally, we found a similar power-law dependence between the magnetic flux $\Phi_V$ and \ro\ (Figure~\ref{fig.b_rossby}b):  {$\Phi_V \propto \ro^{-1.19\pm0.14}$}. Right/left arrows in Figure~\ref{fig.b_rossby} denote the cases with lower/upper limits of \ro.

We note that the correlation between $\bv$ and $\ro$ indeed has less scatter than that between $\bv$ and $P_{\rm rot}$  shown in Figure~\ref{fig.b_prot}.  In spite of the tighter correlation, a noticeable scatter still exists, which, as discussed in Section~\ref{subsec.prot-B}, could be caused by different field topologies. It is also worth noting that the field topology and intensity can change over a stellar magnetic cycle and this fact alone can also be a source of scatter in our relations (although it is possibly not the dominant source).  For the large-scale field of the Sun, a variation of a factor of $\sim 2$ in $\bv$ is observed between the two maps used in this work, when the Sun changed to a simplified, large-scale dipolar topology at solar minimum (CR 1907) from a more complex one at maximum (CR 1851). For stars like HD~190711, the variation of $\bv$ among the maps considered in this study is almost a factor of $3$. 

%%%%%%%%%%%%%%%%%%%
%%%%%%%%%%%%%%%%%%%

\subsubsection{Correlations with X-ray luminosity}
\begin{figure}
\includegraphics[width=80mm]{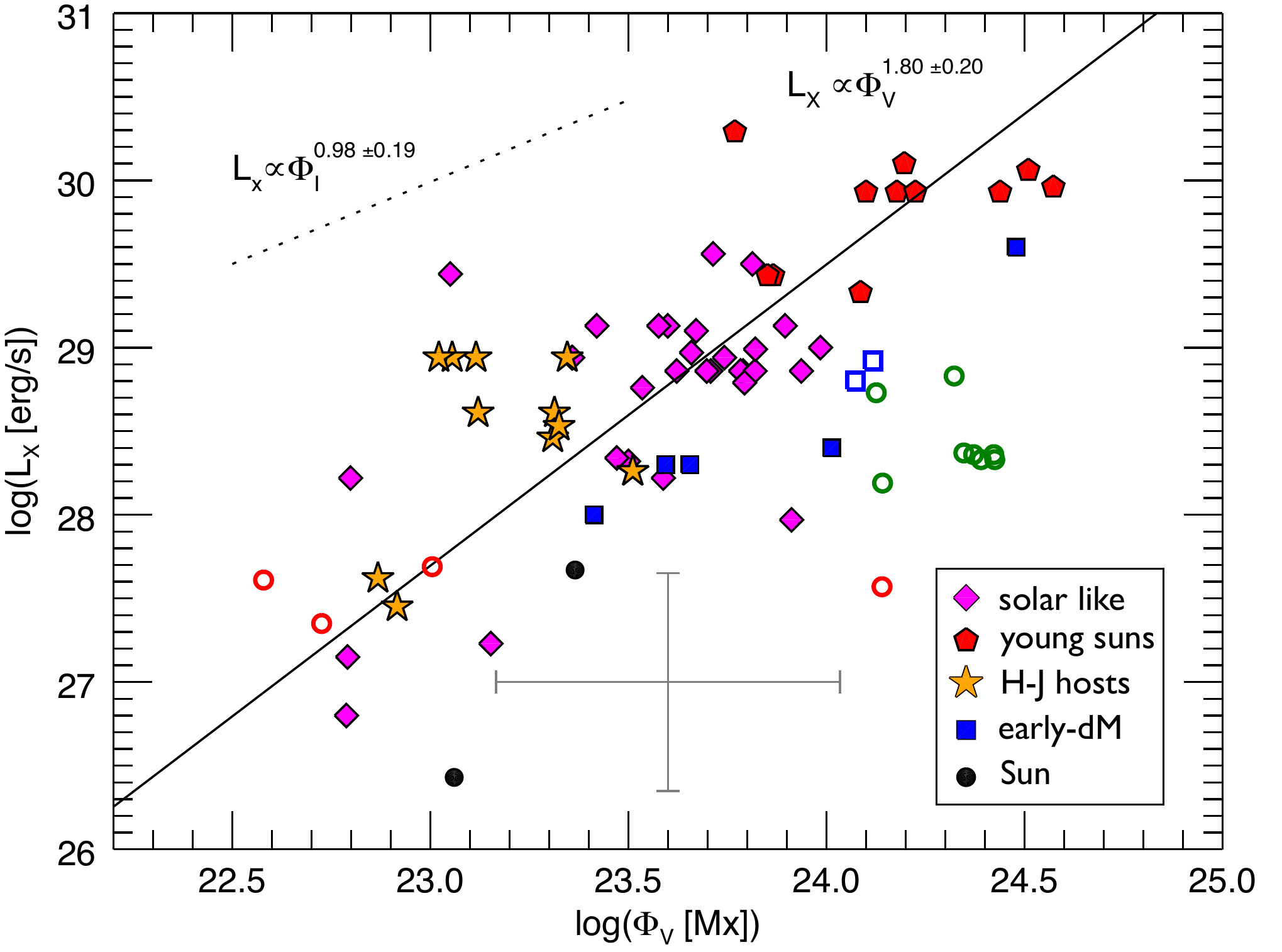}%5_Lx_flux}
\caption{Correlation between X-ray luminosity $L_X$ and large-scale magnetic flux ($\Phi_V=4\pi R_\star^2 \bv$) derived from the ZDI technique for the non-accreting stars in our sample. The open symbols are as in Figure~\ref{fig.b_prot} and were not considered in the fit (solid line). The dotted line, at an arbitrary vertical offset, is indicative of the slope found from ZB measurements for dwarf stars between $L_X$ and $\Phi_I = \bi 4 \pi R_\star^2$ \citep{2003ApJ...598.1387P}. These slopes are  consistent with each other within $3 \sigma$, but samples with a large dynamic range of $\bi$ are desirable to better constrain this result (see text). \label{fig.Lx_flux}}
\end{figure}

Another interesting trend we found in our data is between the X-ray luminosity $L_X$ and $\Phi_V$ (Figure~\ref{fig.Lx_flux}). For the non-accreting stars we found that {$L_X \propto \Phi_V^{1.80 \pm 0.20}$}.  If we include the accreting objects, the slope between $L_X$ and $\Phi_V$ flattens and we find that {$L_X^{\rm (all)} \propto \Phi_V^{0.913 \pm 0.054}$} (fit not shown in Figure~\ref{fig.Lx_flux}).

%%%%%%%%%%%%%%%%%%%
%%%%%%%%%%%%%%%%%%%

We also investigate the trend between the ratio of X-ray-to-bolometric luminosity $L_X/L_{\rm bol}$ and the large-scale magnetic field. Considering the dwarf stars represented by the filled symbols in Figure~\ref{fig.ap2}, we found that {$L_X/L_{\rm bol} \propto \bv^{1.61\pm 0.15}$ (solid line).

%%%%%%%%%%%%%%%%%%%
\begin{figure}
\includegraphics[width=80mm]{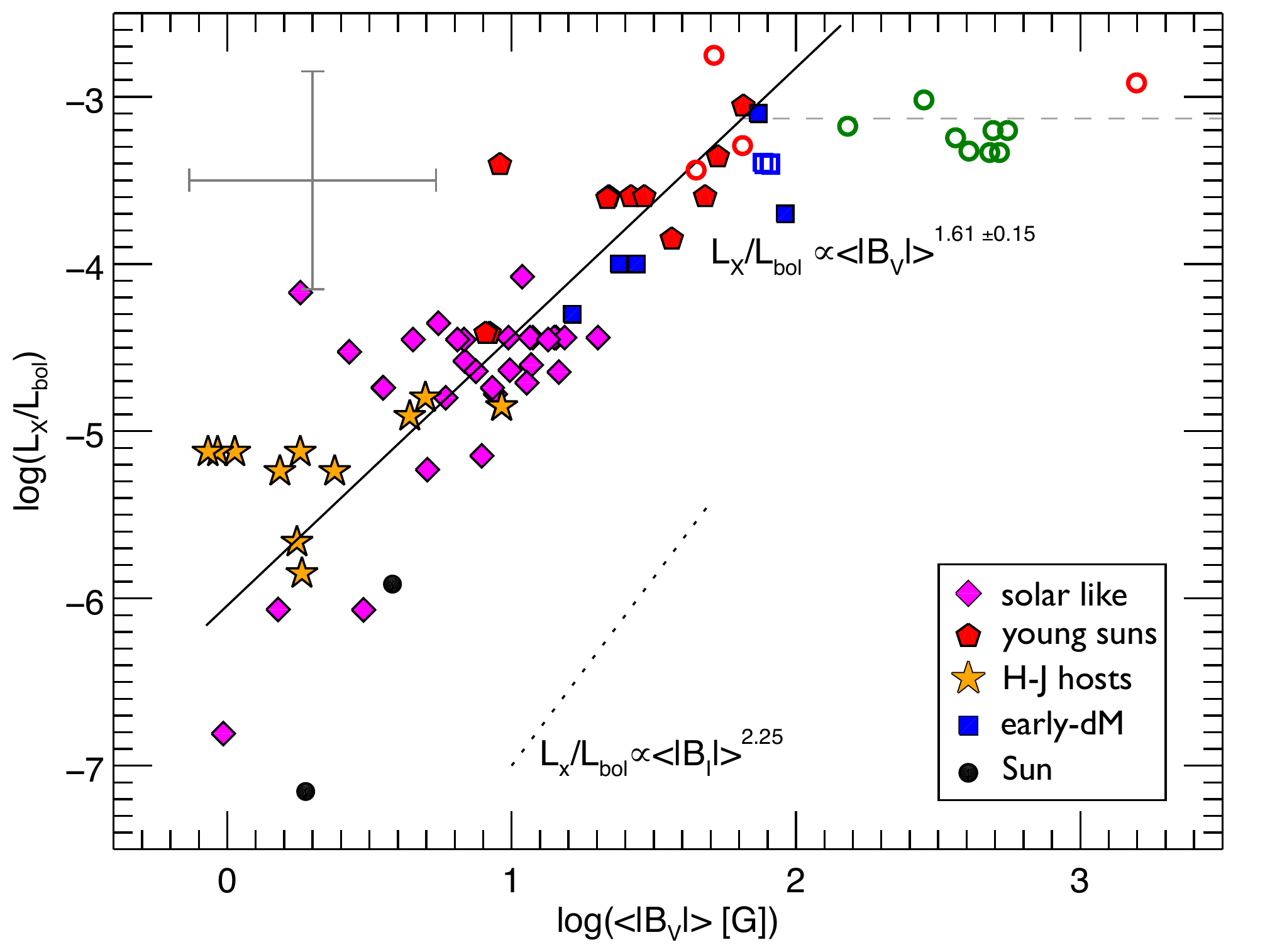}%5_LxLbol_B}
\caption{Correlation between the ratio of X-ray-to-bolometric luminosity ($L_X/L_{\rm bol}$) and large-scale magnetic field  derived from the ZDI technique ($\bv$) for the non-accreting stars in our sample. The open symbols are as in Figure~\ref{fig.b_prot} and were not considered in our fit (solid line). The dashed line indicates the saturation plateau for $\ro \lesssim 0.1$ at $\log(L_X/L_{\rm bol}) \simeq -3.1$  \citep{2011ApJ...743...48W}. The dotted line, at an arbitrary vertical offset, is indicative of the slope found from ZB measurements \citep[derived from results by][]{2001ASPC..223..292S,2011ApJ...743...48W}. 
\label{fig.ap2}}
\end{figure}
%%%%%%%%%%%%%%%%%%%

%%%%%%%%%%%%%%%%%%%%%%%%%%%%%%%%%%%%%%%%%%%
\subsection{Accreting PMS stars}\label{sec.ttauri}
Figure \ref{fig.prot_age} shows that the accreting stars form a different population compared to the disc-less stars. Besides the presence of the disc regulating the rotation of accreting PMS stars, they are also still contracting towards the ZAMS and, therefore, their radii and internal structures evolve considerably over a short timescale (compared to their MS lifetime). 

While the non-accreting stars show significant correlations in almost all the trends investigated in Table~\ref{tab.powerlaw}, the same is not true for the accreting stars. With the exception of the correlation between $\Phi_V$ and $P_{\rm rot}$ (discussed below), all the other trends investigated resulted in relatively low correlation coefficients and/or relatively high null-probabilities ($> 0.01\%$).

In accreting systems,  the polar strength of the dipole component $B_{\rm dip}$ is particularly relevant for determining the disc truncation radius and the balance of torques in the star-disc system \citep[e.g.,][]{2012ApJ...744...55A}. \citet{2012ApJ...755...97G} and, more recently confirmed by \citet{2014MNRAS.437.3202J},  found that $B_{\rm dip}$ is correlated with $P_{\rm rot}$, such that stars with weak dipole components tend to be rotating faster than stars with strong dipole components. They attributed this correlation as a signature of star-disc interaction. Using the data for $B_{\rm dip}$ listed in \citet{2012ApJ...755...97G}, \citet{2013MNRAS.436..881D} and \citet{2014MNRAS.437.3202J} together with the data presented in Table~\ref{table}, we found that $B_{\rm dip} \propto P_{\rm rot}^{2.05\pm 0.41}$, with a Spearman's rank correlation coefficient of $\rho=0.83$ and a probability of the null hypothesis that no correlation exists of $\ll 0.01\%$. In addition, we also found that $\Phi_V \propto B_{\rm dip}^{1.07 \pm0.22}$ ($\rho=0.90$). These two strong correlations directly explain the strong correlation reported in Table \ref{tab.powerlaw} between $\Phi_V$ and $P_{\rm rot}$, where we found that $\Phi_V \propto P_{\rm rot} ^{2.19\pm0.43}$, when the error in $\Phi_V$ is properly accounted for. We caution, however, that these correlations are based on a small sample of accreting stars and more data are required for confirmation.

Accreting PMS stars with the simplest magnetic fields, and the largest magnetic flux, are therefore the slowest rotators.  The correlations reported here are likely a manifestation of the star-disc interaction, as suggested by  \citet{2012ApJ...755...97G}.  Stars with more organised large-scale magnetospheres with stronger dipole components are able to truncate their discs at larger radii, where the Keplerian spin rate of the inner disc (and that of the star if they exist in a disc-locked state) is slower than it would be at the smaller truncation radii expected for stars with more complex magnetospheres with weaker dipole components.  The latter sample of stars, with their lower magnetic flux $\Phi_V$, would therefore be faster rotators.\footnote{If an accreting PMS is not locked to its disc, then a stronger dipole component allows the disc to be truncated at a larger radius, closer to co-rotation.  This in turn means the star will experience smaller magnetic and accretion related spin-up torques (e.g. \citealt{2013arXiv1309.7851B} and references therein), and will more likely remain a slower rotator compared to a star with a weaker more complex magnetic field, as it evolves towards a disc-locked state.}  

Note that, because  most PMS accreting stars observed to date have $\ro \ll 0.1$ and are in the saturated regime, their dynamo-generated magnetic fields are not expected to depend on their rotation rates. The correlation between $\Phi_V$ and $P_{\rm rot}$ we observe is what we would expect if the rotation rates of accreting PMS stars are being dominated by star-disc interaction. In other words, the stellar magnetic field (via star-disc interaction) sets the rotation rate of accreting PMS stars, rather than the rotation rate setting the magnetic flux/strength through the dynamo field generation process.  Or, at the very least, star-disc effects dominate any underlying dynamo relations at this early phase of stellar evolution. 
 
%%%%%%%%%%%%%%%%%%%%%%%%%%%%%%%%%%%%%%%%%%%
\section{Discussion}\label{sec.discussion}
\subsection{Comparison between results from Zeeman broadening and Zeeman-Doppler imaging}\label{sec.comparison}
In this Section, we compare  trends with magnetism. Stellar magnetic fields were obtained by two different techniques. The Zeeman-induced line broadening of unpolarised light (Stokes I), or Zeeman broadening (ZB) technique, yields estimates of the average of the total unsigned surface field strength (small- and large-scale structures), without providing information of the topology of the field. The Zeeman Doppler imaging (ZDI) technique (Stokes V), on the other hand, is able to reconstruct the intensity and topology of the stellar magnetic field, but cannot reconstruct the small-scale field component, which is missed within the resolution element of the reconstructed ZDI maps \citep{2013AN....334...48M}.%
\footnote{The reconstructed fields are expressed as a spherical-harmonic expansion. Note that, the faster the rotation of the star, the larger is the spatial resolution. As a consequence, the ZDI reconstruction technique is able to recover magnetic fluxes at high order $l$ of the spherical harmonics expansion for faster rotating objects \citep[see][for a detailed analysis of the effects of resolution on what is recovered in the ZDI maps]{2009MNRAS.398..189H}. In our sample, the maximum value of $l$ varies from $l_{\rm max} \sim 2 $ \citep[e.g., for HD~76151,][]{2008MNRAS.388...80P} to $\sim 30$ \citep[e.g., for HD~141943,][]{2011MNRAS.413.1922M}. To verify the existence of a possible bias in the reconstructed ZDI field with spatial resolution, we have recalculated $\bv$ for all the objects taking into account only the lowest orders of $l$. We adopted  $l_{\rm cutoff} = \min(5, l_{\rm max})$ and recomputed the power-law indices $p$ for all the relations presented in Table~\ref{tab.powerlaw}. The recalculated $p$ are consistent within the fitting errors to what is presented in Table~\ref{tab.powerlaw}. The similarity between the relations when considering $\bv(l_{\rm cutoff})$ and $\bv(l_{\rm max})$ is due to the fact that the largest powers in the harmonic expansions are in the low-$l$ modes. This indicates that the different spatial resolution of the data considered here does not generate bias in the derived $\bv$ and, consequently, that our derived relations in Table~\ref{tab.powerlaw} are robust. \label{foot}} 

These techniques are, nevertheless, complementary. The ZB technique is limited to slowly rotating objects ($v \sin(i)\lesssim 20$~km~s$^{-1}$), as broadening of spectral lines caused by rotation makes it more difficult to disentangle broadening caused by the Zeeman effect. The ZDI measurements, on the other hand, favour rapidly rotating objects (a few tens of km~s$^{-1}$, although  recently ZDI measurements of more slowly rotating objects have become available). As a result,  it is not always possible to obtain field measurements using both techniques for the same object \citep[see][for a more in depth discussion]{2012EAS....57..165M}. Because of that, in this Section, instead of comparing results of both techniques on a case-by-case basis, we compare the results achieved from these techniques on samples of stars (which in general do not have overlapping members). The comparison presented next is summarised in Table~\ref{tab.comp}. The dotted lines in Figures~\ref{fig.b_prot} to \ref{fig.ap2} indicate the slopes found from ZB measurements, assuming arbitrary vertical offsets.

\begin{table*} 
\centering
\caption{Comparison between trends found using Zeeman Doppler Imaging (ZDI, this work) and Zeeman broadening (ZB) measurements for stars in the unsaturated regime. References for the latter are provided in the last column. \label{tab.comp}}    
\begin{tabular}{lllccc}  
\hline
From ZDI (this work) & From ZB  & Reference\\
\hline
$\langle |B_V| \rangle \propto P_{\rm rot}^{-1.32 \pm 0.14}$ & $\langle |B_I| \rangle \propto P_{\rm rot}^{-1.7}$ &\citet{1996IAUS..176..237S} \\
$\bv \propto \ro^{-1.38 \pm 0.14}$ & $\bi \propto \ro^{-1.2}$ &\citet{2001ASPC..223..292S} \\
$L_X^{\rm (all)} \propto \Phi_V^{0.913 \pm 0.054}$ & $L_X^{\rm (all)} \propto \Phi_I^{1.13}$ &\citet{2003ApJ...598.1387P} \\
$L_X^{\rm (dwarfs)} \propto \Phi_V^{1.80\pm 0.20}$ & $L_X^{\rm (dwarfs)} \propto \Phi_I^{0.98\pm 0.19}$ &\citet{2003ApJ...598.1387P} \\
$L_X/L_{\rm bol} \propto \bv^{1.61 \pm 0.15}$ & $L_X/L_{\rm bol} \propto \bi^{2.25}$ &\citet{2001ASPC..223..292S,2011ApJ...743...48W} \\
\hline
\end{tabular}
\end{table*}
 
Observations of  magnetic fields of about a dozen stars using ZB have revealed that $\langle |B_I| \rangle \propto P_{\rm rot}^{-1.7}$ \citep{1996IAUS..176..237S} and, in terms of Rossby numbers, $\bi \propto \ro^{-1.2}$ \citep{2001ASPC..223..292S}. {In both works, a mix of saturated and unsaturated stars are considered, which implies that if one were to only consider the stars in the unsaturated regime, the slopes would be steeper than the ones  derived by \citet{1996IAUS..176..237S,2001ASPC..223..292S}.} Using the ZDI measurements of the large-scale field $\bv$, we found for the non-accreting stars that {$\langle |B_V| \rangle \propto P_{\rm rot}^{-1.32 \pm 0.14}$ and $\bv \propto \ro^{-1.38 \pm 0.14}$} (the latter considering only points with $\ro \gtrsim 0.1$, corresponding to the unsaturated stars). The similarities in the dependences of $\bi$ and $\bv$ with $P_{\rm rot}$ and $\ro$ might indicate that fields measured by ZDI (large scale) and ZB (large and small scale) are coupled to each other \citep[see also][]{2014MNRAS.439.2122L}. This apparent coupling, therefore, might indicate that small- and large-scale fields share the same dynamo field generation processes, at least for stars in the unsaturated regime.

Another relevant comparison is the one between X-ray emission and magnetism as derived by ZB and ZDI (Figure~\ref{fig.Lx_flux}). \citet{2003ApJ...598.1387P} found that $L_X^{\rm (all)} \propto \Phi_I^{1.13\pm 0.05}$, where $\Phi_I = \bi 4 \pi R_\star^2$ is the unsigned magnetic flux derived from ZB. In this relation, \citet{2003ApJ...598.1387P} considered magnetic field observations of the Sun (quiet Sun, X-ray bright points, active regions, and integrated solar disk), dwarf stars and PMS accreting stars, spanning about $12$ orders of magnitude in magnetic flux. When we include all the objects in our sample, we found that {$L_X^{\rm (all)} \propto \Phi_V^{0.913 \pm 0.054}$}, consistent to the nearly linear trend found by \citet{2003ApJ...598.1387P}.  When considering only the sample of $16$ G, K and M dwarf stars (i.e., no solar data nor accreting PMS stars), \citet{2003ApJ...598.1387P} found that $L_X^{\rm (dwarfs)} \propto \Phi_I^{0.98\pm 0.19}$, which is flatter than the correlation we found ($L_X^{\rm (dwarfs)} \propto \Phi_V^{1.80\pm 0.20}$), based on a larger sample of {$61$} dwarf stars\footnote{Note that if we include the open symbols (M dwarf stars without age estimates) in the fit presented in Figure~\ref{fig.Lx_flux}, the slope we derive is slightly flatter ($L_X^{\rm (dwarfs)} \propto \Phi_V^{1.49\pm 0.17}$), yet still consistent with the value quoted in the text.}. Because of the relatively large errors in the power-law exponent of these relations, within $3 \sigma$ they are still consistent with each other. This is a point worthy of further investigation. Finding a different power law for $\Phi_V$ and $\Phi_I$ may shed light on how the small-scale and large-scale field structures contribute to $L_X$. By reducing the errors in the power-law fits (e.g., increasing the dynamic ranges of the fits, in particular in the ZB one), it would be possible to assess whether these relations are indeed consistent with each other. 

Finally, in Figure~\ref{fig.ap2} we showed that {$L_X/L_{\rm bol} \propto \bv^{1.61\pm 0.15}$} for the unsaturated stars. To the best of our knowledge, there is no such correlation constructed for $\bi$. We therefore combined the results of \citet[][$\bi \propto \ro^{-1.2}$]{2001ASPC..223..292S}  and \citet[][$L_X/L_{\rm bol}\propto \ro^{-2.7 \pm 0.13}$]{2011ApJ...743...48W} to derive that $L_X/L_{\rm bol} \propto \bi^{2.25}$. Again, we note that the slope derived in \citet{2001ASPC..223..292S} could be steeper if only the unsaturated stars were considered. Therefore, the slope of $2.25$ we derive is an upper limit. Although we found  a less steep dependence of $\Rx$ with $\bv$ than with $\bi$, given the uncertainties involved in the determination of these slopes, they can be considered consistent with each other. Unfortunately and in particular because of the small number of unsaturated stars with available $\bi$ measurements, it is still not possible to ascertain how large- and small-scale fields contribute to X-ray emission.

%%%%%%%%%%%%%%%%%%%%%%%%%%%%%%%%%%%%%%
\subsection{Saturation}\label{sec.saturation}
Stars in the saturated regime show similar levels of X-ray-to-bolometric luminosity. In X-rays, saturation occurs for stars with $\ro \lesssim 0.1$ \citep[e.g.,][]{2000MNRAS.318.1217J,2003A_A...397..147P,2011ApJ...743...48W}. In terms of their magnetism, there is evidence that the total field $\bi$ also saturates for $\ro \lesssim 0.1$ \citep{2009ApJ...692..538R} and it would be interesting to investigate whether saturation is also present in the large-scale magnetic field $\bv$. In Figure~\ref{fig.b_rossby}a, we also present the remaining M dwarf stars, without age estimates (open symbols), collected from the samples in \citet{2008MNRAS.390..545D} and \citet{2008MNRAS.390..567M,2010MNRAS.407.2269M}. They are in the X-ray saturated regime, with small $\ro$ (\ro\ taken from \citealt{2008MNRAS.390..545D,2008MNRAS.390..567M,2010MNRAS.407.2269M}). {It seems that these objects show different levels of saturation of $\bv$, with the mid-M dwarfs (green circles) saturating at $\log({\bv}/[{\rm G}]) \sim 2.6$ while the early Ms (blue squares) at $\log({\bv}/[{\rm G}]) \sim 1.7$ (horizontal dashed lines in Figure~\ref{fig.b_rossby}a).  \citet{2008MNRAS.390..545D} suggested that the step in the saturation level between early-Ms and mid-Ms is caused by different efficiencies at producing large-scale versus small-scale fields, where rapidly-rotating mid-M dwarfs generate fields on larger spatial scales than early-M dwarfs \citep[see also][where a direct comparison between $\bv$ and $\bi$ was performed for a small sample of M dwarf stars]{2009A&A...496..787R}. The saturation of late-M dwarfs (red circles), on the other hand, was shown to be divided into two distinct categories, either more similar to the saturation level of early-Ms or that of mid-Ms \citep{2010MNRAS.407.2269M}. Although in \citet{2008MNRAS.390..545D}, \citet{2009A&A...496..787R} and \citet{2010MNRAS.407.2269M} the three components of the reconstructed ZDI field were considered (radial, azimuthal and meridional) and in the present work we only focus on the radial component, the trends obtained in Figure~\ref{fig.b_rossby}a are essentially the same as those discussed by these authors.

A unified saturation plateau for $\ro \lesssim 0.1$ is observed if the magnetic flux $\Phi_V$ is considered instead of the magnetic field intensity $\bv$ (Figure~\ref{fig.b_rossby}b). This occurs at $\log{(\Phi_V/[{\rm Mx}])}\sim 24.25$. There is a spread in this plateau, in particular caused by the late-M dwarfs (red circles). This spread has also been observed in X-rays, for objects later than M$6.5$ \citep{2014ApJ...785...10C}. The saturation of $\Phi_V$ has not been recognised before. Observations of more objects at low $\ro$ are desirable to provide better constraints  on this saturation. }

{In Figure~\ref{fig.ap2}, we investigated how $\Rx$ varied with magnetism. Over-plotted to Figure~\ref{fig.ap2} are the remaining M dwarf stars, without age estimates (open symbols), from the samples in \citet{2008MNRAS.390..545D} and \citet{2008MNRAS.390..567M,2010MNRAS.407.2269M}. The saturation value  of $\log(\Rx) = -3.13\pm 0.08$, derived from the rotation-activity study performed by \citet{2011ApJ...743...48W}, is shown as a dashed line. We see that the mid- and late-M dwarf stars approximately lie along this plateau. We did a similar analysis between  $\Rx$  and magnetic flux $\Phi_V$ and found that in this case, the plateau disappears as early- and mid-M dwarfs lie approximately along the same trend of  $\Rx$ and $\Phi_V$  as the remaining objects ($\Rx \propto \Phi_V^{1.82 \pm 0.18}$).} 

Figure~\ref{fig.sketch} shows a possible interpretation of our results, where we show a three-dimensional sketch of $\Rx$, $\bv$ and \ro. In this sketch, Figures~\ref{fig.b_rossby} ($\bv$ versus \ro) and \ref{fig.ap2} ($\Rx$ versus $\bv$) are projections of a multi-dimensional distribution, as is the well-known relation between $\Rx$ and \ro. These projections are illustrated by dashed lines. According to our interpretation, the saturation plateau is actually a `plane' (grey retangular box), where objects of  different internal structures (i.e., different masses) are located at different regions (drawn as blue stripes in our sketch). Each one of these stripes gives rise to the mass-dependent plateaus in the projected plane of $\{\bv,\ro\}$ (cf.~Figure~\ref{fig.b_rossby}) and it also accounts for the shift in $\bv$ observed for the mid-M dwarfs  in the projected plane of $\{\Rx,\bv\}$ (cf.~Figure~\ref{fig.ap2}). The unsaturated stars  consist of a tighter distribution of points (solid red stripe). In Figure~\ref{fig.sketch}, we place our points in the three-dimensional space of $\{\bv,\ro,\Rx\}$, but it is worth noting that the activity relation is a function of other quantities as well, such as, age and mass.

%%%%%%%%%%%%%%%%%%%
\begin{figure}
\includegraphics[width=80mm]{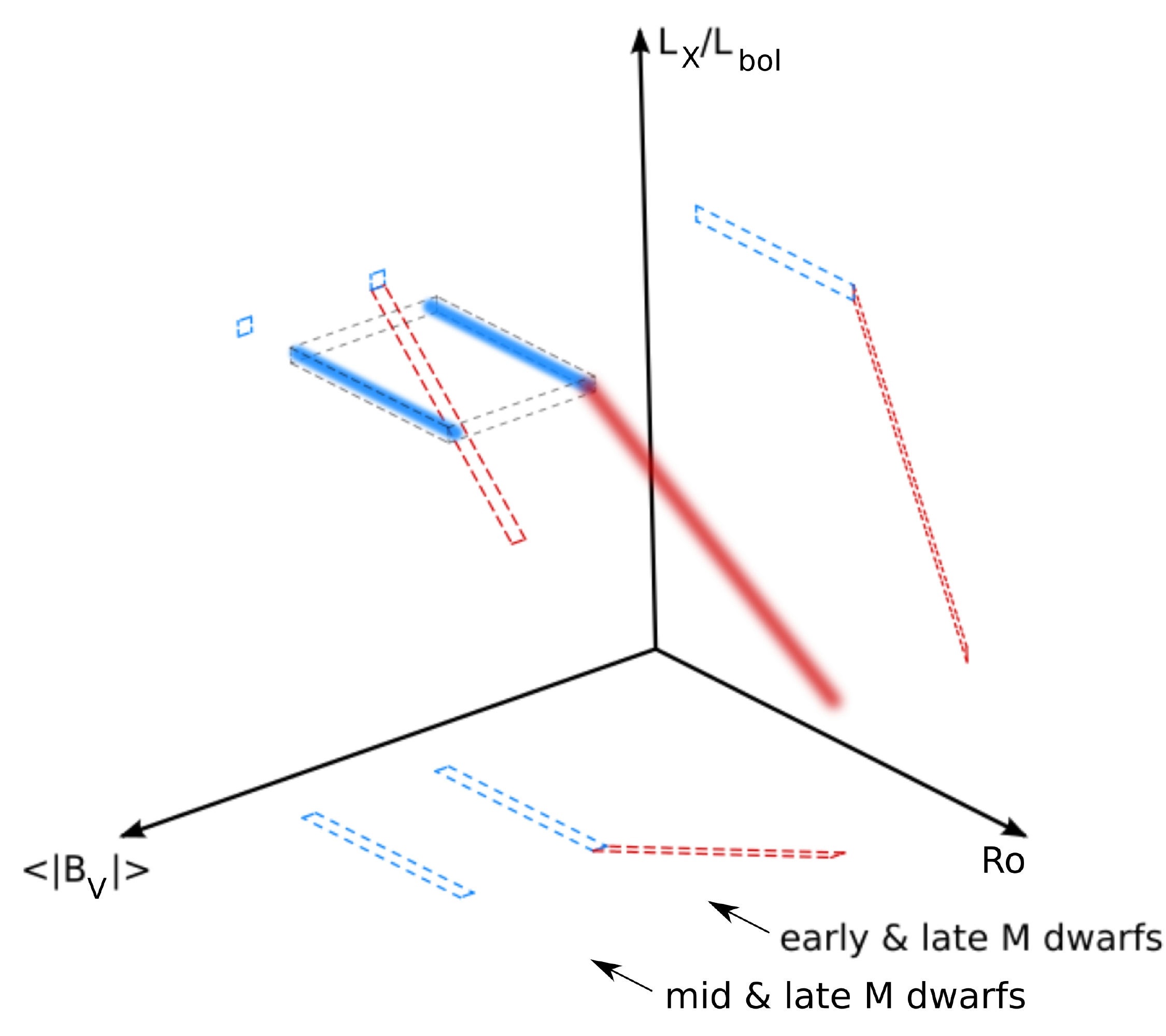}
\caption{The activity relation is a complex function of many variables, such as age, mass, rotation and magnetism. Here we present a sketch in the three-dimensional space of $\{\bv,\ro,\Rx\}$, presenting a possible interpretation of how these quantities are related to each other (blue and red stripes). Figures~\ref{fig.b_rossby} ($\bv$ versus \ro) and \ref{fig.ap2} ($\Rx$ versus $\bv$) are projections of this multi-dimensional distribution, as is the well-known relation between $\Rx$ and \ro. These three projections are illustrated by dashed lines. The saturation plateaus seen in the projections form a saturation `plane' (grey retangular box) in the three-dimensional view, where objects of different masses are located at different regions (blue stripes).
\label{fig.sketch}}
\end{figure}
%%%%%%%%%%%%%%%%%%%

%%%%%%%%%%%%%%%%%%%%%%%%%%%%%%%%%%%%%%%%%%%
\subsection{Stars with hot Jupiters}\label{sec.hjhosts}
Stars with close-in massive planets (or ``hot Jupiters'', hJs) can experience strong tidal forces that may affect their rotation rates. It is believed that some stars that harbour hJs might have spun-up as a consequence of inward planetary migration \citep{1996Natur.380..606L, 2003ApJ...588..509G, 2009MNRAS.396.1789P, 2011MNRAS.415..605B}. Among the stars in our sample, $7$ of them host hJs ($12$ ZDI maps shown as orange symbols in our figures). From Figure~\ref{fig.b_prot}, we note that these stars do not seem to have magnetic and rotation properties that differ from the remaining population of disc-less stars. \citet{2013MNRAS.435.1451F} also compared the large-scale magnetic topology of hJ-host stars adopted in our sample with that of stars without detected hJs and showed that both groups have similar magnetic field topologies. 

Our findings suggest that the planets orbiting the h-J hosts in our sample might not be affecting significantly the rotation nor the large-scale magnetism of their host stars. A possible reason for this might be that tides in the systems analysed here are too weak to spin-up the host star \citep{2010A&A...512A..77L} and, consequently, to change its magnetic properties. Alternatively, if these planets were at some point in the past able to affect the rotation of the star, the reaction of the dynamo should have occurred in a relatively short timescale. 

It is also worth pointing out that the h-J hosts seem to be systematically shifted towards lower $\bv$ values at a given age compared to solar-like stars (Figure~\ref{fig.B_age}). This is likely to be a bias from planet search surveys, which prioritise targets with lower activity and, therefore, lower magnetism.

%%%%%%%%%%%%%%%%%%%%%%%%%%%%%%%%%%%%%%%%%
\subsection{Accreting PMS stars}\label{sec.accreting}
For accreting PMS stars, \citet{2007ApJ...664..975J} found no correlation between any magnetic and stellar/dynamo parameters\footnote{\citet{2011ApJ...729...83Y} also found no correlations between the magnetic and dynamo parameters when considering PMS stars in the Orion Nebula Cluster (ONC), the TW Hydrae association, and the same stars from Taurus considered by \citet{2007ApJ...664..975J}. A comparison of their sample of ONC stars with the catalog of \citet{1998AJ....116.1816H} reveals it to be a mixture of both accreting and non-accreting PMS stars.  However, \citet{2011ApJ...729...83Y} do find a reduction in $\Phi_I$ with age which they attribute to the decrease in stellar radius as PMS contract towards the ZAMS.  We do not find any statistically significant correlation between $\Phi_V$ and age for our population of PMS stars (see Table \ref{tab.powerlaw}).  This may be because our sample size is too small (16 magnetic maps of 11 different accreting PMS stars) compared to the 31 stars considered by \citet{2011ApJ...729...83Y}. Likewise, \citet{2007ApJ...664..975J} found no correlation between $\Phi_I$ and $t$ in his smaller sample of 14 accreting PMS stars.}, and in particular, found no correlation between the magnetic flux $\Phi_I$, estimated from the average surface magnetic field as calculated from ZB measurements, and $P_{\rm rot}$.  Out of the parameters that we have considered in Table \ref{tab.powerlaw}, the only statistically significant correlation we have found for accreting PMS stars is between the magnetic flux $\Phi_V$, derived from the magnetic maps obtained through ZDI, and $P_{\rm rot}$.  As discussed in Section \ref{sec.ttauri}, this is likely being driven by the star-disc interaction, which is controlled by the large-scale field topology probed with ZDI.  ZB studies do not give access to the large-scale field topology, but are sensitive to the entirety of the stellar surface magnetic field, including the small-scale closed field regions that play no part in the star-disc interaction.  The large-scale stellar magnetic field, in particular the dipole component $B_{\rm dip}$ of the multipolar magnetosphere, is the most important in terms of controlling the interaction with the disc (e.g. \citealt{2012ApJ...744...55A,2012ApJ...755...97G,2014MNRAS.437.3202J}); $B_{\rm dip}$ can only be determined from ZDI studies.  Therefore, the lack of correlation between $\Phi_I$ and $P_{\rm rot}$ does not pose a problem for our argument that the clear correlation between $\Phi_V$ and $P_{\rm rot}$ reported in this paper is driven by magnetic star-disc interaction. 

%%%%%%%%%%%%%%%%%%%%%%%%%%%%%%%%%%%%%%%%%%%%
\section{Magnetochronology: magnetism as a new way to assess stellar age}\label{sec.magnetochronology}
One of our most interesting findings is the empirical trend between large-scale magnetism and age. Age is one of the most fundamental stellar parameters. However, the task of measuring ages is a very difficult one, with several methods having been used \citep[see][for recent reviews]{2010ARA&A..48..581S,2013arXiv1311.7024S}. For example, by solving the equations of the internal structure of the star, stellar evolution codes can be used as a tool to determine stellar age, from observational quantities, such as effective temperatures and luminosities. {As in the MS phase, these parameters do not change significantly, isochrone dating is more unreliable for more evolved MS stars.}  
The relation found between $P_{\rm rot}$ and age first recognised by \citetalias{1972ApJ...171..565S} has served as the basis of the gyrochronology method, which is able to provide stellar age estimates from rotation measurements \citep{2003ApJ...586..464B,2010ApJ...721..675B}. For young objects, the presence of lithium can constrain ages \citep{2010IAUS..268..359S}. Asteroseismology can also provide a means to derive stellar age \citep{1988IAUS..123..295C,2005MNRAS.356..671O}, although this method has been more widely applied to bright stars. Chromospheric activity can also be used as an astrophysical clock, although it seems to be more robust for objects with ages $\lesssim 2$~Gyr \citep{2013A&A...551L...8P}. The empirical relation that we identified between the large-scale magnetic fluxes and age (Figure~\ref{fig.B_age}) can be used as an alternative method to estimate the age of stars. However, the relatively large spread of this correlation implies that this method, similarly to other age-dating methods, would carry significant imprecisions in age determination. Moreover, when compared to photometric measurements of rotation periods, the ``magnetochronology'' method is more expensive in terms of observing time and field reconstruction than the gyrochronology method.

Our empirical trends are also relevant for investigations of rotational evolution of low-mass stars, as they provide important constraints on the evolution of the large-scale magnetism of cool stars, as well as their dependence on stellar rotation. For example, the relations $\bv$ versus $t$ and $\bv $ versus $ \ro$ can be implemented in models investigating the evolution of mass and angular momentum loss \citep[e.g.,][]{2013A&A...556A..36G}. These relations also provide important constraints for stellar dynamo studies. 

%%%%%%%%%%%%%%%%%%%%%%%%%%%%%%%%%%%%%%%%%%%
\section{Summary and conclusions}\label{sec.conclusions}
In this paper, we investigated how the large-scale surface magnetic fields of cool dwarf stars, reconstructed using the ZDI technique, vary with age, rotation period, Rossby number and X-ray luminosity. Our sample consists of {$73$} stars in the mass range between {$0.1$ and $2.0~M_\odot$} and spans about 4 orders of magnitude in age (from a Myr to almost $10$~Gyr). As some of the stars have magnetic maps that were obtained at multiple observation epochs, our sample consists of {$104$}~data points, including some PMS objects with on-going accretion. In order to separate the effects that accretion/PMS contraction might play on the rotational evolution of the stars, we have separated our sample into two populations.

For the population of accreting stars, we find few statistically significant correlations, except for the correlation between the unsigned magnetic flux $\Phi_V$ and $P_{\rm rot}$ (and between $\bv$ and the polar strength of the dipole component $B_{\rm dip}$ and $\Phi_V $ versus $ B_{\rm dip}$). We attributed these correlations to a signature of star-disc interaction rather than being caused by the underlying dynamo field generation process.

For the population of non-accreting stars, we showed that their unsigned large-scale magnetic field strength $\langle |B_V| \rangle$ is related to age $t$ as {$\langle |B_V| \rangle \propto t^{-0.655 \pm 0.045}$}, with a high statistical significance (Spearman's rank correlation coefficient of {$-0.79$} and a very small null hypothesis probability). This relation presents a similar power dependence empirically identified in the seminal work of \citetalias{1972ApJ...171..565S}, which has served as the basis of the gyrochronology method to determine stellar ages from stellar rotation measurements. Our empirically-derived magnetism-age relation could be used as a way to estimate stellar ages, although it would not provide better precision than the currently adopted methods. 

Theoretically, \citetalias{1972ApJ...171..565S}'s relation can be explained on the basis of the simplified wind model of \citet{1967ApJ...148..217W}, further assuming that a linear dynamo of the type $B \propto \Omega_\star \propto P_{\rm rot}^{-1}$ is in operation. Empirically, we found that the large-scale unsigned surface field $\langle |B_V| \rangle$ scales with the rotation period of the star as {$\langle |B_V| \rangle \propto P_{\rm rot}^{-1.32 \pm 0.14}$ or, in terms of Rossby number, $\bv \propto \ro^{-1.38\pm 0.14}$}. Our data, therefore, gives support for a linear-type dynamo.  Our empirically-derived relations are relevant for investigations of rotational evolution of low-mass stars and give important observational constrains for stellar dynamo studies. 

We also compared the trends we found in the ZDI data to trends empirically found using Zeeman broadening measurements of magnetic field strengths $\bi$. For the unsaturated stars, the similar dependences of $\bi$ and $\bv$ with $P_{\rm rot}$ and $\ro$ indicates that fields measured by ZDI (large scale) and ZB (large and small scale) are coupled to each other. This might indicate that small- and large-scale fields share the same dynamo field generation processes. For the stars in the saturated regime, saturation of $\bi$ occurs for $\ro \lesssim 0.1$ at {$\bi \sim 3$}~kG \citep[][essentially for M dwarfs]{2009ApJ...692..538R}, while for $\bv$, saturation seems to have a bimodal distribution \citep{2008MNRAS.390..545D} at ${\bv} \sim 10^{1.7}$~G  for the early-Ms and at ${\bv} \sim 10^{2.6}$~G for the mid-Ms. We also found saturation of $\Phi_V$ at  ${\Phi_V} \sim 10^{24.25}$~Mx  for $\ro \lesssim 0.1$, but this is no longer bimodal as in the case of $\bv$. Observations of more objects at low $\ro$ are desirable to provide better constraints  on the saturation of $\Phi_V$. 

We also investigate how the small- and large-scale structures contribute to X-ray emission (Figures~\ref{fig.Lx_flux} and \ref{fig.ap2}). For the unsaturated stars, these contributions between X-ray emission and $\bv$ or $\bi$ have similar slopes within $3\sigma$, but samples with large dynamic range of $\bi$ are required to better constrain this result. 

The plots we presented in this paper could be understood as projections of a complex, multi-dimensional distribution, dependent on quantities such as $\Rx$, $\bv$, rotation, age and internal structure. In Figure~\ref{fig.sketch}, we offered a possible interpretation of this distribution  in the three-dimensional space of $\{\bv,\ro,\Rx\}$. In this view, the unsaturated stars comprise a tight distribution of points, while the saturated objects give rise to a saturation `plane' (instead of a plateau), where objects of different masses are located at different regions (shown as blue stripes in Figure~\ref{fig.sketch}).

New nIR spectropolarimeters, such as SPIRou \citep[e.g.,][]{2013sf2a.conf..497D}, currently under-construction for the Canada-France-Hawaii telescope, will be ideally suited for further comparison between the ZB and ZDI techniques. It will allow magnetically sensitive, Zeeman broadened, lines to be measured within the same spectra as used to reconstruct magnetic maps, thereby allowing a more direct comparison between $\bv$ and $\bi$.

%%%%%%%%%%%%%%%%%%%%%%%%%%%%%%%%%%%%%%%%%%%%%%%%%%
\section*{Acknowledgements}
AAV acknowledges support from a Royal Astronomical Society Fellowship and from the Swiss National Science Foundation via an {\it Ambizione} Fellowship. SGG acknowledges support from the Science \& Technology Facilities Council (STFC) via an Ernest Rutherford Fellowship [ST/J003255/1]. JB, PP, and CPF acknowledge support from the ANR 2011 Blanc SIMI5-6 020 01 ``Toupies: Towards understanding the spin evolution of stars'' (\url{http://ipag.osug.fr/Anr_Toupies/}). AAV would like to thank Prof.~Keith Horne and Dr Kate Rowlands for advice in the statistical analysis. NSO/Kitt Peak data used here are produced cooperatively by NSF/NOAO, NASA/GSFC, and NOAA/SEL.

\def\aj{{AJ}}                   % Astronomical Journal
\def\araa{{ARA\&A}}             % Annual Review of Astron and Astrophys
\def\apj{{ApJ}}                 % Astrophysical Journal
\def\apjl{{ApJ}}                % Astrophysical Journal, Letters
\def\apjs{{ApJS}}               % Astrophysical Journal, Supplement
\def\ao{{Appl.~Opt.}}           % Applied Optics
\def\apss{{Ap\&SS}}             % Astrophysics and Space Science
\def\aap{{A\&A}}                % Astronomy and Astrophysics
\def\aapr{{A\&A~Rev.}}          % Astronomy and Astrophysics Reviews
\def\aaps{{A\&AS}}              % Astronomy and Astrophysics, Supplement
\def\azh{{AZh}}                 % Astronomicheskii Zhurnal
\def\baas{{BAAS}}               % Bulletin of the AAS
\def\jrasc{{JRASC}}             % Journal of the RAS of Canada
\def\memras{{MmRAS}}            % Memoirs of the RAS
\def\mnras{{MNRAS}}             % Monthly Notices of the RAS
\def\pra{{Phys.~Rev.~A}}        % Physical Review A: General Physics
\def\prb{{Phys.~Rev.~B}}        % Physical Review B: Solid State
\def\prc{{Phys.~Rev.~C}}        % Physical Review C
\def\prd{{Phys.~Rev.~D}}        % Physical Review D
\def\pre{{Phys.~Rev.~E}}        % Physical Review E
\def\prl{{Phys.~Rev.~Lett.}}    % Physical Review Letters
\def\pasp{{PASP}}               % Publications of the ASP
\def\pasj{{PASJ}}               % Publications of the ASJ
\def\qjras{{QJRAS}}             % Quarterly Journal of the RAS
\def\skytel{{S\&T}}             % Sky and Telescope
\def\solphys{{Sol.~Phys.}}      % Solar Physics
\def\sovast{{Soviet~Ast.}}      % Soviet Astronomy
\def\ssr{{Space~Sci.~Rev.}}     % Space Science Reviews
\def\zap{{ZAp}}                 % Zeitschrift fuer Astrophysik
\def\nat{{Nature}}              % Nature
\def\iaucirc{{IAU~Circ.}}       % IAU Cirulars
\def\aplett{{Astrophys.~Lett.}} % Astrophysics Letters
\def\apspr{{Astrophys.~Space~Phys.~Res.}}   % Astrophysics Space Physics Research
\def\bain{{Bull.~Astron.~Inst.~Netherlands}}    % Bulletin Astronomical Institute of the Netherlands
\def\fcp{{Fund.~Cosmic~Phys.}}  % Fundamental Cosmic Physics
\def\gca{{Geochim.~Cosmochim.~Acta}}        % Geochimica Cosmochimica Acta
\def\grl{{Geophys.~Res.~Lett.}} % Geophysics Research Letters
\def\jcp{{J.~Chem.~Phys.}}      % Journal of Chemical Physics
\def\jgr{{J.~Geophys.~Res.}}    % Journal of Geophysics Research
\def\jqsrt{{J.~Quant.~Spec.~Radiat.~Transf.}}   % Journal of Quantitiative Spectroscopy and Radiative Transfer
\def\memsai{{Mem.~Soc.~Astron.~Italiana}}   % Mem. Societa Astronomica Italiana
\def\nphysa{{Nucl.~Phys.~A}}    % Nuclear Physics A
\def\physrep{{Phys.~Rep.}}      % Physics Reports
\def\physscr{{Phys.~Scr}}       % Physica Scripta
\def\planss{{Planet.~Space~Sci.}}           % Planetary Space Science
\def\procspie{{Proc.~SPIE}}     % Proceedings of the SPIE
\def\actaa{{Acta~Astronomica}}     % Acta Astronomica
\def\pasa{{Publications of the ASA}}     % Publications of the ASA
\def\na{{New Astronomy}}     % New Astronomy

\let\astap=\aap
\let\apjlett=\apjl
\let\apjsupp=\apjs
\let\applopt=\ao
\let\mnrasl=\mnras

\bsp

%%%%%%%%%%%%%%%%%%%%%%%%%%%%%%
\appendix
\section{Error estimates}\label{ap.sample}
In our fitting procedures, measurement errors were always accounted for. Typical error bars are indicated in the plots presented in this paper (grey error bars). In this Appendix, we describe how the errors in the quantities plotted in this paper were estimated. 

%%%%%%%%%%%%%%%%%%
\subsection{Ages}
The ages we adopted in this paper are listed in Table~\ref{table}. They were compiled from different works in the literature and were derived by different methods. 
Although some of the ages of our stars are reasonably well-constrained (e.g., some of our stars are members of associations and open clusters), most of them do not have assigned errors. In this paper, we have adopted a conservative error estimate of $0.434$~dex in $\log{t}$ for all the stars in our sample. This is equivalent to adopting $\sigma_t=t$ and accounts for the fact that the ages of older stars are in general more poorly constrained than the ages of younger ones. 

%%%%%%%%%%%%%%%%%%
\subsection {Magnetic field measurements}
In the present work, the unsigned surface magnetic field strength $\bv$ and flux $\Phi_V$ are calculated based on the radial component of the observed surface field. We have adopted in this paper a conservative error of $\sigma_{\Phi_V}=\Phi_V$ and $\sigma_{\bv}=\bv$. This results in an error of about $0.434$~dex in $\log (\bv)$ and $\log (\Phi_V)$. Note that in the derivation of magnetic fluxes, we have not taken into consideration errors in the radii of stars. 

{We have also verified the effects of the spatial resolution on the field recovered by the ZDI technique, by artificially restricting the spherical harmonic expansion to low orders. We showed that the different spatial resolution of the data considered here does not generate bias in the derived $\bv$ and $\Phi_V$, and, consequently, that our derived relations are robust. More details of this analysis are provided in footnote \ref{foot}.}

%%%%%%%%%%%%%%%%%%
\subsection {X-ray luminosities}
Because of coronal variability, it is likely that the values of $L_X$ presented in Table \ref{table} are not the same as one would have derived if X-ray observations were to occur simultaneously with spectropolarimetric ones. For the Sun, it is observed that during its activity cycle, the X-ray luminosity varies from $\simeq 0.27$ to $4.7\times 10^{27}$~erg~s$^{-1}$ at minimum and maximum phases, respectively \citep{2000ApJ...528..537P}. This represents a variation of about $90\%$ from an average luminosity between these two extremes. Likewise, it is expected that stars also show X-ray variability during their cycles. To account for possible variations in $L_X$ over stellar cycles, we have assigned an error of $0.651$~dex in $\log L_X$ for all the objects in our sample, which is equivalent as assuming $\sigma_{L_X}=1.5L_X$.

%%%%%%%%%%%%%%%%%%
\subsection {Rotation periods}
Rotation periods are usually well constrained in the literature. In light of that and that errors are significantly larger for ages, magnetic fields and X-ray luminosities, we have neglected errors in rotation periods.

%%%%%%%%%%%%%%%%%%
\subsection {Rossby numbers}\label{ap.rossby}
In the literature, Rossby numbers \ro\ are usually preferred over rotation periods as they allow comparison across different spectral types, yielding tighter correlations (e.g., compare Figures~\ref{fig.b_prot} and \ref{fig.b_rossby}). In this work, we did not assign errors to the computed \ro , but we caution that, to compute \ro, one needs to know the convective turnover time $\tau_c$. To produce Figure~\ref{fig.b_rossby} and the results shown in Table~\ref{tab.powerlaw}, we adopted $\tau_c$ from \citet{2010A&A...510A..46L}. Because we used the same model to compute \ro\ for all our non-accreting stars, these data points should have similar systematic errors.

However, we remind the reader that $\tau_c$ and, consequently, \ro\ are  model-dependent quantities. To investigate the robustness of our relations against \ro\ for the non-accreting stars, we calculated \ro\ using  two other different approaches. In the first approach, we interpolated from $\tau_c$ listed in \citet{2010ApJ...721..675B}, derived for an age of $500$~Myr. As the internal structure of the star does not change significantly after it has entered in the MS phase, $\tau_c$ should not change considerably, such that values listed by \citet{2010ApJ...721..675B} can still provide a reasonable estimate of \ro. In the second approach, we computed \ro\ using the empirical $\tau_c$--$M_\star$ relation found by \citet{2011ApJ...743...48W}. 

Table~\ref{tab.rossby} summarises the power-law indices found when \ro\ was computed using $\tau_c$ from the models of  \citet[LMV2010]{2010A&A...510A..46L} and \citet[BK2010]{2010ApJ...721..675B} and the empirically-derived relation from \citet[W2011]{2011ApJ...743...48W}. The power-law indices derived from the theoretical models (LMV2010 and BK2010) are essentially identical within $1 \sigma$. Comparing these indices with the ones derived using the empirical determination of \ro\ (W2011), we again found reasonably good agreement (within $2\sigma$). This shows that the relations we found against \ro\ are robust and, overall, are not significantly affected by the method adopted to derive $\tau_c$.

\begin{table} 
\centering
\caption{Power-law indices $p$ ($Y\propto X^p$) computed by linear least-squares fit to logarithms for Rossby numbers calculated using different approaches: LMV2010 use the theoretical derivation of $\tau_c$ from \citet{2010A&A...510A..46L}, BK2010 from \citet{2010ApJ...721..675B} and W2011 use the empirical derivation of $\tau_c$ from \citet{2011ApJ...743...48W}.  The fits only consider non-accreting F, G, K, M dwarf stars. In spite of the use of different relations to compute $\tau_c$, all the fits are consistent with each other within $2\sigma$. %\hll{updated 10/feb/14}
\label{tab.rossby}}    
\begin{tabular}{lccccc}  
\hline
$Y$ & $X$& LMV2010 & BK2010 & W2011 \\
\hline
$\bv$&$\ro$   & $ -1.130 \pm 0.087$ & $ -1.051 \pm 0.084$& $ -1.41 \pm 0.12$ \\
$\Phi_V$&$\ro$ & $ -1.143 \pm 0.083$ & $ -0.952 \pm 0.095$& $ -1.31 \pm 0.11$ \\
\hline
\end{tabular}
\end{table}

\label{lastpage}

\end{document}

%% file: latex_table_a.tex
 \\ 
\hline \multicolumn{12}{c}{Solar like stars} \\
HD   3651&K0V & $  0.88$ & $  0.88$ & $43.4$ & $1.916$ & $8200$ & $   27.23$ & $
   -6.07$ & $3.01$ & $-$&
{\bf \citetalias{Petit2014}},\citetalias{2013arXiv1311.3374M},\citetalias{2011A_A...530A..73C}
 \\
HD   9986&G5V & $  1.02$ & $  1.04$ & $23.0$ & $1.621$ & $4300$ & $ - $ & $ - 
$ & $0.517$ & $-$&{\bf \citetalias{Petit2014}},\citetalias{2013arXiv1311.3374M}
 \\
HD  10476&K1V & $  0.82$ & $  0.82$ & $16.0$ & $0.576$ & $8700$ & $   27.15$ & $
   -6.07$ & $1.51$ & $-$&
{\bf \citetalias{Petit2014}},\citetalias{2013arXiv1311.3374M},\citetalias{2006ARep...50..579K}
 \\
HD  20630&G5Vv & $  1.03$ & $  0.95$ & $9.30$ & $0.593$ & $600$ & $   28.79$ & $
   -4.71$ & $11.3$ & 2012 Oct&
\citetalias{2013arXiv1310.7620D},\citetalias{2013arXiv1311.3374M},{\bf \citetalias{2010ApJ...714..384R}},\citetalias{2012ApJ...753...76W}
 \\
HD  22049&K2Vk & $  0.86$ & $  0.77$ & $10.3$ & $0.366$ & $440$ & $   28.32$ & $
   -4.78$ & $8.76$ & $-$&
\citetalias{Petit2014},\citetalias{2013arXiv1311.3374M},{\bf \citetalias{2008A_A...488..771J}},\citetalias{2012ApJ...753...76W}
 \\
HD  39587&G0VCH & $  1.03$ & $  1.05$ & $4.83$ & $0.295$ & $500$ & $   28.99
$ & $   -4.64$ & $9.85$ & $-$&
\citetalias{Petit2014},\citetalias{2013arXiv1311.3374M},{\bf \citetalias{2003AJ....125.1980K}},\citetalias{2012ApJ...753...76W}
 \\
HD  56124&G0 & $  1.03$ & $  1.01$ & $18.0$ & $1.307$ & $4500$ & $   29.44$ & $
   -4.17$ & $1.81$ & $-$&
{\bf \citetalias{Petit2014}},\citetalias{2013arXiv1311.3374M},\citetalias{2011ApJ...743...48W}
 \\
HD  72905&G1.5Vb & $  1.00$ & $  1.00$ & $5.00$ & $0.272$ & $500$ & $   28.97
$ & $   -4.64$ & $7.49$ & $-$&
\citetalias{Petit2014},\citetalias{2013arXiv1311.3374M},{\bf \citetalias{2003AJ....125.1980K}},\citetalias{2012ApJ...753...76W}
 \\
HD  73350&G5V & $  1.04$ & $  0.98$ & $12.3$ & $0.777$ & $510$ & $   28.76$ & $
   -4.80$ & $5.86$ & $-$&
\citetalias{Petit2014},\citetalias{2013arXiv1311.3374M},{\bf \citetalias{2009ApJ...698.1068P}}
 \\
HD  75332&F7Vn & $  1.21$ & $  1.24$ & $4.80$ & $>1.105$ & $1800$ & $   29.56
$ & $   -4.35$ & $5.52$ & $-$&
{\bf \citetalias{Petit2014}},\citetalias{2013arXiv1311.3374M},\citetalias{2007Ap.....50..187B}
 \\
HD  76151&G3V & $  1.24$ & $  0.98$ & $20.5$ & $ - $ & $3600$ & $   28.34$ & $
   -5.23$ & $5.05$ & 2007 Jan&
{\bf \citetalias{2008MNRAS.388...80P}},\citetalias{2003A_A...397..147P} \\
HD  78366&F9V & $  1.34$ & $  1.03$ & $11.4$ & $>2.781$ & $2500$ & $   28.94
$ & $   -4.74$ & $3.54$ & 2011&
\citetalias{2011AN....332..866M},{\bf \citetalias{Petit2014}},\citetalias{2006ARep...50..579K}
 \\
\ldots&\ldots & $\ldots$ & $\ldots$ & $\ldots$ & $\ldots$ & $\ldots$ & $\ldots
$ & $\ldots$ & $8.55$ & 2008&
\citetalias{2011AN....332..866M},{\bf \citetalias{Petit2014}},\citetalias{2006ARep...50..579K}
 \\
\ldots&\ldots & $\ldots$ & $\ldots$ & $\ldots$ & $\ldots$ & $\ldots$ & $\ldots
$ & $\ldots$ & $3.52$ & 2010&
\citetalias{2011AN....332..866M},{\bf \citetalias{Petit2014}},\citetalias{2006ARep...50..579K}
 \\
HD 101501&G8V & $  0.85$ & $  0.90$ & $17.6$ & $0.663$ & $5100$ & $   28.22$ & $
   -5.15$ & $7.85$ & $-$&
{\bf \citetalias{Petit2014}},\citetalias{2013arXiv1311.3374M},\citetalias{2012NewA...17..537X}
 \\
HD 131156A&G8V & $  0.93$ & $  0.84$ & $5.56$ & $0.256$ & $2000$ & $   28.86
$ & $   -4.44$ & $11.9$ & 2010 Jan&
\citetalias{2012A_A...540A.138M},\citetalias{2013arXiv1311.3374M},{\bf \citetalias{Petit2014}},\citetalias{2012ApJ...753...76W}
 \\
\ldots&\ldots & $\ldots$ & $\ldots$ & $\ldots$ & $\ldots$ & $\ldots$ & $\ldots
$ & $\ldots$ & $14.3$ & 2009 Jun&
\citetalias{2012A_A...540A.138M},\citetalias{2013arXiv1311.3374M},{\bf \citetalias{Petit2014}},\citetalias{2012ApJ...753...76W}
 \\
\ldots&\ldots & $\ldots$ & $\ldots$ & $\ldots$ & $\ldots$ & $\ldots$ & $\ldots
$ & $\ldots$ & $11.6$ & 2010 Aug&
\citetalias{2012A_A...540A.138M},\citetalias{2013arXiv1311.3374M},{\bf \citetalias{Petit2014}},\citetalias{2012ApJ...753...76W}
 \\
\ldots&\ldots & $\ldots$ & $\ldots$ & $\ldots$ & $\ldots$ & $\ldots$ & $\ldots
$ & $\ldots$ & $15.4$ & 2010 Jun&
\citetalias{2012A_A...540A.138M},\citetalias{2013arXiv1311.3374M},{\bf \citetalias{Petit2014}},\citetalias{2012ApJ...753...76W}
 \\
\ldots&\ldots & $\ldots$ & $\ldots$ & $\ldots$ & $\ldots$ & $\ldots$ & $\ldots
$ & $\ldots$ & $14.1$ & 2011 Jan&
\citetalias{2012A_A...540A.138M},\citetalias{2013arXiv1311.3374M},{\bf \citetalias{Petit2014}},\citetalias{2012ApJ...753...76W}
 \\
\ldots&\ldots & $\ldots$ & $\ldots$ & $\ldots$ & $\ldots$ & $\ldots$ & $\ldots
$ & $\ldots$ & $9.74$ & 2008 Feb&
\citetalias{2012A_A...540A.138M},\citetalias{2013arXiv1311.3374M},{\bf \citetalias{Petit2014}},\citetalias{2012ApJ...753...76W}
 \\
\ldots&\ldots & $\ldots$ & $\ldots$ & $\ldots$ & $\ldots$ & $\ldots$ & $\ldots
$ & $\ldots$ & $20.1$ & 2007 Aug&
\citetalias{2012A_A...540A.138M},\citetalias{2013arXiv1311.3374M},{\bf \citetalias{Petit2014}},\citetalias{2012ApJ...753...76W}
 \\
HD 131156B&K4V & $  0.99$ & $  1.07$ & $10.3$ & $0.611$ & $2000$ & $   27.97
$ & $   -4.60$ & $11.7$ & $-$&
{\bf \citetalias{Petit2014}},\citetalias{2013arXiv1311.3374M},\citetalias{2012ApJ...753...76W}
 \\
HD 146233&G2V & $  0.98$ & $  1.02$ & $22.7$ & $1.324$ & $4700$ & $   26.80$ & $
   -6.81$ & $0.969$ & 2007 Aug&
{\bf \citetalias{2008MNRAS.388...80P}},\citetalias{2009IAUS..258..395G} \\
HD 166435&G1IV & $  1.04$ & $  0.99$ & $3.43$ & $0.259$ & $3800$ & $   29.50
$ & $   -4.08$ & $10.9$ & $-$&
{\bf \citetalias{Petit2014}},\citetalias{2013arXiv1311.3374M},\citetalias{2001A_A...379..279Q}
 \\
HD 175726&G5 & $  1.06$ & $  1.06$ & $3.92$ & $0.272$ & $500$ & $   29.10$ & $
   -4.58$ & $6.85$ & $-$&
\citetalias{Petit2014},\citetalias{2013arXiv1311.3374M},{\bf \citetalias{2009A_A...501..941H}},\citetalias{2003A_A...397..987C}
 \\
HD 190771&G5IV & $  0.96$ & $  0.98$ & $8.80$ & $0.453$ & $2700$ & $   29.13
$ & $   -4.45$ & $13.4$ & 2010&
\citetalias{2011AN....332..866M},{\bf \citetalias{2008MNRAS.388...80P}},\citetalias{2004A_A...417..651S}
 \\
\ldots&\ldots & $\ldots$ & $\ldots$ & $\ldots$ & $\ldots$ & $\ldots$ & $\ldots
$ & $\ldots$ & $6.45$ & 2009&
\citetalias{2009A_A...508L...9P},{\bf \citetalias{2008MNRAS.388...80P}},\citetalias{2004A_A...417..651S}
 \\
\ldots&\ldots & $\ldots$ & $\ldots$ & $\ldots$ & $\ldots$ & $\ldots$ & $\ldots
$ & $\ldots$ & $4.50$ & 2008&
\citetalias{2009A_A...508L...9P},{\bf \citetalias{2008MNRAS.388...80P}},\citetalias{2004A_A...417..651S}
 \\
\ldots&\ldots & $\ldots$ & $\ldots$ & $\ldots$ & $\ldots$ & $\ldots$ & $\ldots
$ & $\ldots$ & $6.80$ & 2007&
{\bf \citetalias{2008MNRAS.388...80P}},\citetalias{2004A_A...417..651S} \\
HD 201091A&K5V & $  0.66$ & $  0.62$ & $34.2$ & $0.786$ & $3600$ & $   28.22
$ & $   -4.53$ & $2.68$ & $-$&
\citetalias{Petit2014},\citetalias{2013arXiv1311.3374M},\citetalias{2002ApJ...574..412W},{\bf \citetalias{2008ApJ...687.1264M}}
 \\
HD 206860&G0V & $  1.10$ & $  1.04$ & $4.55$ & $0.388$ & $260$ & $   29.00$ & $
   -4.65$ & $14.7$ & $-$&
\citetalias{Petit2014},\citetalias{2013arXiv1311.3374M},{\bf \citetalias{2013A_A...556A..53E}},\citetalias{2003A_A...397..147P}
 \\
\hline \multicolumn{12}{c}{Young suns} \\
BD-16 351&K5 & $  0.90$ & $  0.83$ & $3.39$ & $ - $ & $30$ & $ - $ & $ - $ & $
33.4$ & 2012 Sep&\citetalias{Folsom2014},{\bf \citetalias{2008hsf2.book..757T}}
 \\
HD  29615&G3V & $  0.95$ & $  0.96$ & $2.32$ & $0.073$ & $27$ & $ - $ & $ - 
$ & $45.1$ & 2009&
\citetalias{Waite2014},\citetalias{2011PASA...28..323W},\citetalias{2011A_A...532A..10M},{\bf \citetalias{2008hsf2.book..757T},\citetalias{2008ApJ...689.1127M}}
 \\
HD  35296&F8V & $  1.22$ & $  1.20$ & $3.90$ & $>0.467$ & $35$ & $   29.43$ & $
   -4.41$ & $8.37$ & 2007 Jan&
{\bf \citetalias{Waite2014}},\citetalias{2013arXiv1311.3374M},\citetalias{1995A_A...302..775G}
 \\
\ldots&\ldots & $\ldots$ & $\ldots$ & $\ldots$ & $\ldots$ & $\ldots$ & $\ldots
$ & $\ldots$ & $8.10$ & 2008 Jan&
{\bf \citetalias{Waite2014}},\citetalias{2013arXiv1311.3374M},\citetalias{1995A_A...302..775G}
 \\
HD  36705&K1V & $  1.00$ & $  1.00$ & $0.510$ & $0.028$ & $120$ & $   30.06$ & $
   -3.36$ & $53.1$ & 1996&
\citetalias{1999MNRAS.302..437D},\citetalias{2009A_ARv..17..251S},\citetalias{2011MNRAS.410.2472A},{\bf \citetalias{2013ApJ...766....6B},\citetalias{2005ApJ...628L..69L}},\citetalias{2012ApJ...753...76W}
 \\
HD 106506&G1V & $  1.50$ & $  2.15$ & $1.39$ & $>0.024$ & $10$ & $ - $ & $ - 
$ & $30.8$ & 2007 Apr&{\bf \citetalias{2011MNRAS.413.1949W}} \\
HD 129333&G1.5V & $  1.04$ & $  0.97$ & $2.77$ & $0.177$ & $120$ & $   29.93
$ & $   -3.60$ & $47.9$ & 2012 Jan&
\citetalias{Waite2014},\citetalias{2013arXiv1311.3374M},{\bf \citetalias{2013ApJ...766....6B},\citetalias{2005ApJ...628L..69L}},\citetalias{2012ApJ...753...76W}
 \\
\ldots&\ldots & $\ldots$ & $\ldots$ & $\ldots$ & $\ldots$ & $\ldots$ & $\ldots
$ & $\ldots$ & $22.0$ & 2007 Jan&
\citetalias{Waite2014},\citetalias{2013arXiv1311.3374M},{\bf \citetalias{2013ApJ...766....6B},\citetalias{2005ApJ...628L..69L}},\citetalias{2012ApJ...753...76W}
 \\
\ldots&\ldots & $\ldots$ & $\ldots$ & $\ldots$ & $\ldots$ & $\ldots$ & $\ldots
$ & $\ldots$ & $29.3$ & 2007 Feb&
\citetalias{Waite2014},\citetalias{2013arXiv1311.3374M},{\bf \citetalias{2013ApJ...766....6B},\citetalias{2005ApJ...628L..69L}},\citetalias{2012ApJ...753...76W}
 \\
\ldots&\ldots & $\ldots$ & $\ldots$ & $\ldots$ & $\ldots$ & $\ldots$ & $\ldots
$ & $\ldots$ & $26.3$ & 2006 Dec&
\citetalias{Waite2014},\citetalias{2013arXiv1311.3374M},{\bf \citetalias{2013ApJ...766....6B},\citetalias{2005ApJ...628L..69L}},\citetalias{2012ApJ...753...76W}
 \\
HD 141943&G2V & $  1.30$ & $  1.60$ & $2.18$ & $>0.085$ & $17$ & $ - $ & $ - 
$ & $27.8$ & 2009 Apr&{\bf \citetalias{2011MNRAS.413.1922M}} \\
\ldots&\ldots & $\ldots$ & $\ldots$ & $\ldots$ & $\ldots$ & $\ldots$ & $\ldots
$ & $\ldots$ & $45.8$ & 2007 Mar&{\bf \citetalias{2011MNRAS.413.1922M}} \\
\ldots&\ldots & $\ldots$ & $\ldots$ & $\ldots$ & $\ldots$ & $\ldots$ & $\ldots
$ & $\ldots$ & $36.1$ & 2010 Mar&{\bf \citetalias{2011MNRAS.413.1922M}} \\
HD 171488&G0V & $  1.06$ & $  1.09$ & $1.31$ & $0.089$ & $40$ & $   30.10$ & $
   -3.61$ & $21.7$ & 2004 Sep&
{\bf \citetalias{2006MNRAS.370..468M}},\citetalias{2003A_A...411..595S},\citetalias{2003A_A...399..983W}
 \\
HII 296        &K3 & $  0.80$ & $  0.74$ & $2.61$ & $ - $ & $130$ & $   29.33
$ & $   -3.85$ & $36.6$ & 2009 Oct&
\citetalias{Folsom2014},{\bf \citetalias{1998ApJ...499L.199S}},\citetalias{2003A_A...410..671M}
 \\
HII 739        &G3 & $  1.08$ & $  1.03$ & $2.70$ & $ - $ & $130$ & $   30.29
$ & $   -3.41$ & $9.09$ & 2009 Oct&
\citetalias{Folsom2014},{\bf \citetalias{1998ApJ...499L.199S}},\citetalias{2003A_A...410..671M}
 \\
HIP 12545      &K6 & $  0.58$ & $  0.57$ & $4.83$ & $ - $ & $21$ & $ - $ & $ - 
$ & $78.5$ & 2012 Sep&
\citetalias{Folsom2014},{\bf \citetalias{2014MNRAS.438L..11B}} \\
HIP 76768      &K6 & $  0.61$ & $  0.60$ & $3.64$ & $ - $ & $120$ & $ - $ & $ - 
$ & $54.2$ & 2013 May&
\citetalias{Folsom2014},{\bf \citetalias{2013ApJ...766....6B}},{\bf \citetalias{2005ApJ...628L..69L}}
 \\
LQ Hya&K2V & $  0.80$ & $  0.97$ & $1.60$ & $0.053$ & $50$ & $   29.96$ & $
   -3.06$ & $65.3$ & 1998 Dec&
\citetalias{2003MNRAS.345.1145D},\citetalias{2004A_A...417.1047K},{\bf \citetalias{2004ApJ...614..386B}}
 \\
TYC0486-4943-1&K3 & $  0.69$ & $  0.68$ & $3.75$ & $ - $ & $120$ & $ - $ & $ - 
$ & $20.1$ & 2013 Jun&
\citetalias{Folsom2014},{\bf \citetalias{2013ApJ...766....6B}},{\bf \citetalias{2005ApJ...628L..69L}}
 \\
TYC5164-567-1 &K2 & $  0.85$ & $  0.79$ & $4.71$ & $ - $ & $120$ & $ - $ & $ - 
$ & $39.4$ & 2013 Jun&
\citetalias{Folsom2014},{\bf \citetalias{2013ApJ...766....6B}},{\bf \citetalias{2005ApJ...628L..69L}}
 \\
TYC6349-0200-1&K6 & $  0.54$ & $  0.54$ & $3.39$ & $ - $ & $21$ & $ - $ & $ - 
$ & $34.1$ & 2013 Jun&
\citetalias{Folsom2014},{\bf \citetalias{2014MNRAS.438L..11B}} \\
TYC6878-0195-1&K4 & $  0.65$ & $  0.64$ & $5.72$ & $ - $ & $21$ & $ - $ & $ - 
$ & $31.7$ & 2013 Jun&
\citetalias{Folsom2014},{\bf \citetalias{2014MNRAS.438L..11B}} \\

%% file: latex_table_b.tex
 \\ 
\hline \multicolumn{12}{c}{Hot-Jupiter hosts} \\
$\tau$ Boo&F7V & $  1.34$ & $  1.42$ & $3.00$ & $>0.732$ & $2500$ & $   28.94
$ & $   -5.12$ & $1.06$ & 2008 Jan&
\citetalias{2009MNRAS.398.1383F},\citetalias{2013MNRAS.435.1451F},{\bf \citetalias{2005A_A...443..609S}},\citetalias{2010A_A...515A..98P}
 \\
\ldots&\ldots & $\ldots$ & $\ldots$ & $\ldots$ & $\ldots$ & $\ldots$ & $\ldots
$ & $\ldots$ & $1.81$ & 2007 Jun&
\citetalias{2008MNRAS.385.1179D},\citetalias{2013MNRAS.435.1451F},{\bf \citetalias{2005A_A...443..609S}},\citetalias{2010A_A...515A..98P}
 \\
\ldots&\ldots & $\ldots$ & $\ldots$ & $\ldots$ & $\ldots$ & $\ldots$ & $\ldots
$ & $\ldots$ & $0.856$ & 2006 Jun&
\citetalias{2007MNRAS.374L..42C},\citetalias{2013MNRAS.435.1451F},{\bf \citetalias{2005A_A...443..609S}},\citetalias{2010A_A...515A..98P}
 \\
\ldots&\ldots & $\ldots$ & $\ldots$ & $\ldots$ & $\ldots$ & $\ldots$ & $\ldots
$ & $\ldots$ & $0.925$ & 2008 Jul&
\citetalias{2009MNRAS.398.1383F},\citetalias{2013MNRAS.435.1451F},{\bf \citetalias{2005A_A...443..609S}},\citetalias{2010A_A...515A..98P}
 \\
HD  46375&K1IV & $  0.97$ & $  0.86$ & $42.0$ & $2.340$ & $5000$ & $   27.45
$ & $   -5.85$ & $1.83$ & 2008 Jan&
\citetalias{2013MNRAS.435.1451F},{\bf \citetalias{2005A_A...443..609S}},\citetalias{2008ApJ...687.1339K}
 \\
HD  73256&G8 & $  1.05$ & $  0.89$ & $14.0$ & $0.962$ & $830$ & $   28.53$ & $
   -4.91$ & $4.38$ & 2008 Jan&
\citetalias{2013MNRAS.435.1451F},{\bf \citetalias{2003A_A...407..679U}},\citetalias{2008ApJ...687.1339K}
 \\
HD 102195&K0V & $  0.87$ & $  0.82$ & $12.3$ & $0.473$ & $2400$ & $   28.46$ & $
   -4.80$ & $4.98$ & 2008 Jan&
\citetalias{2013MNRAS.435.1451F},{\bf \citetalias{2006ApJ...648..683G}},\citetalias{2010A_A...515A..98P}
 \\
HD 130322&K0V & $  0.79$ & $  0.83$ & $26.1$ & $0.782$ & $930$ & $   27.62$ & $
   -5.66$ & $1.76$ & 2008 Jan&
\citetalias{2013MNRAS.435.1451F},{\bf \citetalias{2005A_A...443..609S}},\citetalias{2010A_A...515A..98P}
 \\
HD 179949&F8V & $  1.21$ & $  1.19$ & $7.60$ & $>1.726$ & $2100$ & $   28.61
$ & $   -5.24$ & $1.53$ & 2007 Jun&
\citetalias{2012MNRAS.423.1006F},\citetalias{2013MNRAS.435.1451F},{\bf \citetalias{2005A_A...443..609S}},\citetalias{2010A_A...515A..98P}
 \\
\ldots&\ldots & $\ldots$ & $\ldots$ & $\ldots$ & $\ldots$ & $\ldots$ & $\ldots
$ & $\ldots$ & $2.39$ & 2009 Sep&
\citetalias{2012MNRAS.423.1006F},\citetalias{2013MNRAS.435.1451F},{\bf \citetalias{2005A_A...443..609S}},\citetalias{2010A_A...515A..98P}
 \\
HD 189733&K2V & $  0.82$ & $  0.76$ & $12.5$ & $0.403$ & $600$ & $   28.26$ & $
   -4.85$ & $9.21$ & 2008 Jul&
\citetalias{2010MNRAS.406..409F},\citetalias{2013MNRAS.435.1451F},{\bf \citetalias{2006A_A...460..251M}},\citetalias{2010A_A...515A..98P}
 \\
\ldots&\ldots & $\ldots$ & $\ldots$ & $\ldots$ & $\ldots$ & $\ldots$ & $\ldots
$ & $\ldots$ & $9.23$ & 2007 Jun&
\citetalias{2010MNRAS.406..409F},\citetalias{2013MNRAS.435.1451F},{\bf \citetalias{2006A_A...460..251M}},\citetalias{2010A_A...515A..98P}
 \\
\hline \multicolumn{12}{c}{M dwarf stars} \\
CE Boo&M2.5 & $  0.48$ & $  0.43$ & $14.7$ & $<0.288$ & $130$ & $   28.40$ & $
   -3.70$ & $91.6$ & 2008 Jan&
\citetalias{2008MNRAS.390..545D},{\bf \citetalias{1998ApJ...499L.199S}} \\
DS Leo&M0 & $  0.58$ & $  0.52$ & $14.0$ & $<0.267$ & $710$ & $   28.30$ & $
   -4.00$ & $23.9$ & 2007 Dec&
\citetalias{2008MNRAS.390..545D},{\bf \citetalias{2014MNRAS.438.1162V}} \\
\ldots&\ldots & $\ldots$ & $\ldots$ & $\ldots$ & $\ldots$ & $\ldots$ & $\ldots
$ & $\ldots$ & $27.4$ & 2007 Jan&
\citetalias{2008MNRAS.390..545D},{\bf \citetalias{2014MNRAS.438.1162V}} \\
GJ 182&M0.5 & $  0.75$ & $  0.82$ & $4.35$ & $0.054$ & $21$ & $   29.60$ & $
   -3.10$ & $73.6$ & 2007 Jan&
\citetalias{2008MNRAS.390..545D},{\bf \citetalias{2014MNRAS.438L..11B}} \\
GJ 49&M1.5 & $  0.57$ & $  0.51$ & $18.6$ & $<0.352$ & $1200$ & $   28.00$ & $
   -4.30$ & $16.3$ & 2007 Jul&
\citetalias{2008MNRAS.390..545D},{\bf \citetalias{2014MNRAS.438.1162V}} \\
AD Leo&M3 & $  0.42$ & $  0.38$ & $2.24$ & $0.047$ & $ - $ & $   28.73$ & $
   -3.18$ & $152$ & 2008 Feb&\citetalias{2008MNRAS.390..567M} \\
DT Vir&M0.5 & $  0.59$ & $  0.53$ & $2.85$ & $0.092$ & $ - $ & $   28.92$ & $
   -3.40$ & $76.6$ & 2008 Feb&\citetalias{2008MNRAS.390..545D} \\
EQ Peg A&M3.5 & $  0.39$ & $  0.35$ & $1.06$ & $0.020$ & $ - $ & $   28.83$ & $
   -3.02$ & $282$ & 2006 Aug&\citetalias{2008MNRAS.390..567M} \\
EQ Peg B&M4.5 & $  0.25$ & $  0.25$ & $0.400$ & $0.005$ & $ - $ & $   28.19$ & $
   -3.25$ & $364$ & 2006 Aug&\citetalias{2008MNRAS.390..567M} \\
EV Lac&M3.5 & $  0.32$ & $  0.30$ & $4.37$ & $0.068$ & $ - $ & $   28.37$ & $
   -3.32$ & $406$ & 2007 Aug&\citetalias{2008MNRAS.390..567M} \\
GJ 1111&M6 & $  0.10$ & $  0.11$ & $0.460$ & $0.005$ & $ - $ & $   27.61$ & $
   -2.75$ & $51.5$ & 2009&
\citetalias{2010MNRAS.407.2269M},\citetalias{1995ApJ...450..392S} \\
GJ 1156&M5 & $  0.14$ & $  0.16$ & $0.490$ & $0.005$ & $ - $ & $   27.69$ & $
   -3.29$ & $64.9$ & 2009&
\citetalias{2010MNRAS.407.2269M},\citetalias{2011ApJ...743...48W} \\
GJ 1245B&M5.5 & $  0.12$ & $  0.14$ & $0.710$ & $0.007$ & $ - $ & $   27.35$ & $
   -3.44$ & $44.5$ & 2008&
\citetalias{2010MNRAS.407.2269M},\citetalias{2011ApJ...743...48W} \\
OT Ser&M1.5 & $  0.55$ & $  0.49$ & $3.40$ & $0.097$ & $ - $ & $   28.80$ & $
   -3.40$ & $81.0$ & 2008 Feb&\citetalias{2008MNRAS.390..545D} \\
V374 Peg&M4 & $  0.28$ & $  0.28$ & $0.450$ & $0.006$ & $ - $ & $   28.36$ & $
   -3.20$ & $493$ & 2006 Aug&
\citetalias{2008MNRAS.384...77M},\citetalias{2008MNRAS.390..567M} \\
\ldots&\ldots & $\ldots$ & $\ldots$ & $\ldots$ & $\ldots$ & $\ldots$ & $\ldots
$ & $\ldots$ & $554$ & 2005 Aug&
\citetalias{2008MNRAS.384...77M},\citetalias{2008MNRAS.390..567M} \\
WX UMa&M6 & $  0.10$ & $  0.12$ & $0.780$ & $0.008$ & $ - $ & $   27.57$ & $
   -2.92$ & $1580$ & 2009&
\citetalias{2010MNRAS.407.2269M},\citetalias{1995ApJ...450..392S} \\
YZ CMi&M4.5 & $  0.32$ & $  0.29$ & $2.77$ & $0.042$ & $ - $ & $   28.33$ & $
   -3.33$ & $520$ & 2007 Feb&\citetalias{2008MNRAS.390..567M} \\
\ldots&\ldots & $\ldots$ & $\ldots$ & $\ldots$ & $\ldots$ & $\ldots$ & $\ldots
$ & $\ldots$ & $480$ & 2008 Feb&\citetalias{2008MNRAS.390..567M} \\
\hline \multicolumn{12}{c}{Sun} \\
Max [CR1851]&G2V & $  1.00$ & $  1.00$ & $25.0$ & $1.577$ & $4600$ & $   27.67
$ & $   -5.91$ & $3.81$ & 1982 Jan&
\citetalias{2000ApJ...528..537P}, {\bf \citetalias{2010NatGe...3..637B}} \\
Min [CR1907]&G2V & $  1.00$ & $  1.00$ & $25.0$ & $1.577$ & $4600$ & $   26.43
$ & $   -7.15$ & $1.89$ & 1986 Mar&
\citetalias{2000ApJ...528..537P}, {\bf \citetalias{2010NatGe...3..637B}} \\
\hline \multicolumn{12}{c}{Classical T Tauri stars} \\
AA Tau&K7 & $  0.70$ & $  2.00$ & $8.22$ & $0.036$ & $1.4$ & $   30.08$ & $
   -3.50$ & $918$ & 2009 Jan&
\citetalias{2010MNRAS.409.1347D},\citetalias{2014MNRAS.437.3202J},{\bf \citetalias{2000A_A...358..593S}},{\bf \citetalias{2012ApJ...755...97G}},\citetalias{2007A_A...468..353G}
 \\
BP Tau&K7 & $  0.70$ & $  1.95$ & $7.60$ & $0.032$ & $1.9$ & $   30.15$ & $
   -3.40$ & $685$ & 2006 Feb&
\citetalias{2008MNRAS.386.1234D},\citetalias{2014MNRAS.437.3202J},{\bf \citetalias{2000A_A...358..593S}},{\bf \citetalias{2012ApJ...755...97G}},\citetalias{2007A_A...468..353G}
 \\
\ldots&\ldots & $\ldots$ & $\ldots$ & $\ldots$ & $\ldots$ & $\ldots$ & $\ldots
$ & $\ldots$ & $654$ & 2006 Dec&
\citetalias{2008MNRAS.386.1234D},\citetalias{2014MNRAS.437.3202J},{\bf \citetalias{2000A_A...358..593S}},{\bf \citetalias{2012ApJ...755...97G}},\citetalias{2007A_A...468..353G}
 \\
CR Cha&K2  & $  1.90$ & $  2.50$ & $2.30$ & $0.025$ & $2.8$ & $   30.30$ & $
   -3.86$ & $161$ & 2006 Apr&
\citetalias{2009MNRAS.398..189H},\citetalias{2014MNRAS.437.3202J},{\bf \citetalias{2000A_A...358..593S}},{\bf \citetalias{2012ApJ...755...97G}},\citetalias{2011AJ....141..127I}
 \\
CV Cha&G8  & $  2.00$ & $  2.50$ & $4.40$ & $0.079$ & $4.8$ & $   30.11$ & $
   -4.36$ & $170$ & 2006 Apr&
\citetalias{2009MNRAS.398..189H},\citetalias{2014MNRAS.437.3202J},{\bf \citetalias{2000A_A...358..593S}},{\bf \citetalias{2012ApJ...755...97G}},\citetalias{1993ApJ...416..623F}
 \\
DN Tau&M0 & $  0.65$ & $  1.90$ & $6.32$ & $0.027$ & $1.7$ & $   30.08$ & $
   -3.41$ & $195$ & 2012 Dec&
{\bf \citetalias{2000A_A...358..593S}},{\bf \citetalias{2013MNRAS.436..881D}},\citetalias{2007A_A...468..353G}
 \\
\ldots&\ldots & $\ldots$ & $\ldots$ & $\ldots$ & $\ldots$ & $\ldots$ & $\ldots
$ & $\ldots$ & $317$ & 2010 Dec&
{\bf \citetalias{2000A_A...358..593S}},{\bf \citetalias{2013MNRAS.436..881D}},\citetalias{2007A_A...468..353G}
 \\
GQ Lup&K7 & $  1.05$ & $  1.70$ & $8.40$ & $0.042$ & $3.4$ & $   29.87$ & $
   -3.71$ & $600$ & 2011 Jun&
\citetalias{2012MNRAS.425.2948D},\citetalias{2014MNRAS.437.3202J},{\bf \citetalias{2000A_A...358..593S}},{\bf \citetalias{2012ApJ...755...97G}},\citetalias{2010A_A...519A.113G}
 \\
\ldots&\ldots & $\ldots$ & $\ldots$ & $\ldots$ & $\ldots$ & $\ldots$ & $\ldots
$ & $\ldots$ & $761$ & 2009 Jul&
\citetalias{2012MNRAS.425.2948D},\citetalias{2014MNRAS.437.3202J},{\bf \citetalias{2000A_A...358..593S}},{\bf \citetalias{2012ApJ...755...97G}},\citetalias{2010A_A...519A.113G}
 \\
TW Hya&K7 & $  0.80$ & $  1.10$ & $3.56$ & $0.020$ & $9.6$ & $   30.32$ & $
   -2.80$ & $885$ & 2008 Mar&
\citetalias{2011MNRAS.417..472D},\citetalias{2014MNRAS.437.3202J},{\bf \citetalias{2000A_A...358..593S}},{\bf \citetalias{2012ApJ...755...97G}},\citetalias{2010A_A...519A.113G}
 \\
\ldots&\ldots & $\ldots$ & $\ldots$ & $\ldots$ & $\ldots$ & $\ldots$ & $\ldots
$ & $\ldots$ & $1120$ & 2010 Mar&
\citetalias{2011MNRAS.417..472D},\citetalias{2014MNRAS.437.3202J},{\bf \citetalias{2000A_A...358..593S}},{\bf \citetalias{2012ApJ...755...97G}},\citetalias{2010A_A...519A.113G}
 \\
V2129 Oph&K5  & $  1.35$ & $  2.00$ & $6.53$ & $0.036$ & $3.7$ & $   30.43$ & $
   -3.30$ & $499$ & 2005 Jun&
\citetalias{2011MNRAS.412.2454D},\citetalias{2014MNRAS.437.3202J},{\bf \citetalias{2000A_A...358..593S}},{\bf \citetalias{2012ApJ...755...97G}},\citetalias{2011A_A...530A...1A}
 \\
\ldots&\ldots & $\ldots$ & $\ldots$ & $\ldots$ & $\ldots$ & $\ldots$ & $\ldots
$ & $\ldots$ & $644$ & 2009 Jul&
\citetalias{2011MNRAS.412.2454D},\citetalias{2014MNRAS.437.3202J},{\bf \citetalias{2000A_A...358..593S}},{\bf \citetalias{2012ApJ...755...97G}},\citetalias{2011A_A...530A...1A}
 \\
V2247 Oph&M1  & $  0.36$ & $  2.00$ & $3.50$ & $0.016$ & $1.4$ & $   30.11$ & $
   -3.14$ & $142$ & 2008 Jul&
\citetalias{2010MNRAS.402.1426D},\citetalias{2014MNRAS.437.3202J},{\bf \citetalias{2000A_A...358..593S}},{\bf \citetalias{2012ApJ...755...97G}},\citetalias{2010A_A...519A..34P}
 \\
V4046 Sgr A&K5  & $  0.95$ & $  1.12$ & $2.42$ & $0.021$ & $16$ & $   30.08$ & $
   -3.11$ & $69.1$ & 2009 Sep&
\citetalias{2011MNRAS.417.1747D},\citetalias{2014MNRAS.437.3202J},{\bf \citetalias{2000A_A...358..593S}},{\bf \citetalias{2012ApJ...755...97G}},\citetalias{2012ApJ...747..142S}
 \\
V4046 Sgr B&K5 & $  0.85$ & $  1.04$ & $2.42$ & $0.019$ & $17$ & $   30.08$ & $
   -2.93$ & $102$ & 2009 Sep&
\citetalias{2011MNRAS.417.1747D},\citetalias{2014MNRAS.437.3202J},{\bf \citetalias{2000A_A...358..593S}},{\bf \citetalias{2012ApJ...755...97G}},\citetalias{2012ApJ...747..142S}
 \\

%% file: fit_bisector_table_for_paper.tex
t & $P_{\rm rot}$ &
$0.76$ & $< 0.01 $ & $ 1.96 \pm 0.15$ 
 & 
$-0.42$ & $20$ & $ -1.68 \pm 0.59$ 
&
 $0.66$ & $< 0.01 $ & $ 2.54 \pm 0.19$ 
 \\ 
$\bv$ & t &
$-0.79$ & $< 0.01 $ & $ -0.655 \pm 0.045$ 
 & 
$-0.12$ & $65$ & $ -1.03 \pm 0.42$ 
&
 $-0.87$ & $< 0.01 $ & $ -0.701 \pm 0.028$ 
 \\ 
 $\Phi_V$ & t &
$-0.81$ & $< 0.01 $ & $ -0.622 \pm 0.042$ 
 & 
$-0.33$ & $21$ & $ -1.26 \pm 0.35$ 
&
 $-0.89$ & $< 0.01 $ & $ -0.840 \pm 0.029$ 
 \\ 
$\bv$ & $P_{\rm rot}$ &
$-0.54$ & $< 0.01 $ & $ -1.32 \pm 0.14$ 
 & 
$0.61$ & $1.3$ & $ 1.78 \pm 0.49$ 
&
 $-0.44$ & $< 0.01 $ & $ -1.72 \pm 0.17$
 \\ 
%\ldots & \ldots & $-0.50^a$ & $<0.01^a$ & $-1.36 \pm 0.17^a$\\
$\Phi_V$ & $P_{\rm rot}$ &
$-0.72$ & $< 0.01 $ & $ -1.31 \pm 0.11$ 
 & 
$0.82$ & $< 0.01 $ & $ 2.19 \pm 0.43$ 
&
 $-0.57$ & $< 0.01 $ & $ -2.06 \pm 0.18$ 
 \\ 
  %\ldots & \ldots & $-0.69^a$ & $<0.01^a$ & $-1.37 \pm 0.13^a$\\
$\bv$ & Ro &
%$-0.86$ & $< 0.01 $ & $ -1.130 \pm 0.087$  valor original, para todos os Ro
$-0.80^a$ & $<0.01^a$ & $-1.38 \pm 0.14^a$
 & 
$0.27$ & $32$ & $ 1.48 \pm 0.81$ 
&
 $-0.91$ & $< 0.01 $ & $ -1.325 \pm 0.058$ 
 \\ 
%\ldots & \ldots & $-0.80^a$ & $<0.01^a$ & $-1.38 \pm 0.14^a$\\
$\Phi_V$ & Ro &
%$-0.81$ & $< 0.01 $ & $ -1.143 \pm 0.083$  valor original, para todos os Ro
$-0.71^a$ & $<0.01^a$ & $-1.19 \pm 0.14^a$ 
 & 
$0.59$ & $1.5$ & $ 2.30 \pm 0.74$ 
&
 $-0.88$ & $< 0.01 $ & $ -1.596 \pm 0.065$ 
 \\ 
%\ldots & \ldots & $-0.71^a$ & $<0.01^a$ & $-1.19 \pm 0.14^a$\\

$L_X$ & $\Phi_V$ &
$0.64$ & $< 0.01 $ & $ 1.80 \pm 0.20$ 
 & 
$0.20$ & $46$ & $ 0.70 \pm 0.50$ 
&
 $0.80$ & $< 0.01 $ & $ 0.913 \pm 0.054$ 
 \\ 
$L_X/L_{\rm bol}$ & $\langle |B_V| \rangle$ &
$0.81$ & $< 0.01 $ & $ 1.61 \pm 0.15$ 
 & 
$0.059$ & $83$ & $ 1.01 \pm 0.52$ 
&
 $0.87$ & $< 0.01 $ & $ 1.071 \pm 0.067$ 
 \\ 
$L_X/L_{\rm bol}$ & $\Phi_V$ &
$0.79$ & $< 0.01 $ & $ 1.82 \pm 0.18$ 
 & 
$-0.23$ & $38$ & $ -0.92 \pm 0.38$ 
 & 
$0.85$ & $< 0.01 $ & $ 0.894 \pm 0.055$ \\